\documentclass[aps, pre, twocolumn, groupedaddress]{revtex4-1}

\makeatletter

\usepackage{amsmath,amssymb,amsfonts,bm,braket}
\usepackage{graphicx}
\usepackage{subfigure}
\renewcommand{\vec}[1]{\bm{#1}}
\newcommand{\mat}[1]{\mathbf{#1}}

\DeclareMathOperator{\re}{Re}
\DeclareMathOperator{\im}{Im}
\DeclareMathOperator{\diag}{diag}
\DeclareMathOperator{\adj}{adj}

\DeclareMathOperator*{\argsup}{argsup}

\makeatother

\begin{document}

\preprint{APS/123-QED}

\title{Spatio-temporal linear instability analysis for arbitrary dispersion relations on the Lefschetz thimble in multidimensional spacetime}

\author{Taiki Morinaga}

\affiliation{Graduate School of Advanced Science and Engineering, Waseda University, 3-4-1 Okubo, Shinjuku, Tokyo 169-8555, Japan.}

\author{Shoichi Yamada}

\affiliation{Graduate School of Advanced Science and Engineering, Waseda University, 3-4-1 Okubo, Shinjuku, Tokyo 169-8555, Japan.}

\date{\today}
\begin{abstract}
In linear stability analysis of field quantities described by partial differential equations, the well-established classical theory is all but impossible to apply to concrete problems in its entirety even for uniform backgrounds when the spatial dimension is more than 1. In this study, using the Lefschetz thimble method, we develop a new formalism to give an explicit expression to the asymptotic behavior of linear perturbations. It is not only more mathematically rigorous than the previous theory but also useful practically in its applications to realistic problems, and, as such, has an impact on broad subjects in physics.
\end{abstract}
\maketitle

\section{Introduction}
In many dynamical problems of complex nonlinear systems, it is often important to investigate the perturbative dynamics near fixed points, or time-independent solutions, by linearization, since the original nonlinear equations are too difficult to treat directly. For systems with a finite degree of freedom, it suffices to obtain normal-mode solutions with an exponential time-dependence, $\exp(- i \omega t)$, and see if there is a complex $\omega$ with a positive imaginary part. In the case of infinite degrees of freedom, however, this may not be the case. In particular, the linear analysis of the dynamics of field quantities defined in an unlimited spatial domain is much more involved, since the single normal mode has a zero-measure and we need to study the behavior of wave packets; the concept of ``instability'' itself needs to be extended a bit: solutions that grow exponentially in time at every spatial point, i.e., unstable in the ordinary sense, are called ``absolutely unstable'' whereas those solutions that are damped in time at each fixed point in space but are amplified exponentially if seen from an appropriate moving observer are referred to as ``convectively unstable.'' It is then insufficient to see just the sign of the imaginary parts of frequency $\omega(\vec{k})$, this time a function of wave number $\vec{k}$, for each normal mode.

All these things are well known, e.g., in plasma physics and a mathematically rigorous theory was established by Briggs \cite{Briggs1964} and others already in 1960's. It is given in some standard text books \cite{Landau1997}. In this theory, the asymptotic behavior of a perturbation given as a wave packet is derived by finding a coalescence of some of the complex roots of the dispersion relation (DR) $\Delta(\omega,k)=0$. Since they treated only the case with one spatial dimension, the wave number is not a vector but a number actually. They employed the Laplace transform and its inverse to obtain a solution in the form of complex integrals over $\omega$ and $k$. The dominant contribution to these integrals at  asymptotically large times comes from a pole in the upper-hemisphere of the complex $\omega$-plane, which is in turn produced by the coalescence of two roots in $k$ of the DR and the resultant pinch of the integration contour in $k$ at this complex value of $\omega$. The point, at which two complex roots in $k$ of the DR merge, is referred to as the critical point and in addition to $\Delta (\omega, k) = 0,\ \partial \Delta (\omega, k) / \partial k = 0$ is also satisfied there. Note that not all the critical points give the pinch of the integration contour. The theory has been applied to shear flows \cite{Huerre1985,Huerre1990}, jets \cite{Guillot2007}, solitons \cite{Kamchatnov2008} and even neutrino oscillations recently \cite{Capozzi2017, Yi2019}. 

Brevdo extended this theory to 2 spatial dimensions in 1991 \cite{Brevdo1991}. Although it is indeed a straightforward extension of the Briggs' theory, which just looks for root mergers repeatedly, carrying it out is actually much more difficult than in the one spatial dimension and it is essentially impossible to apply unless the DR is very simple. It is not difficult to imagine then that it is hopeless to extend the theory to higher dimensions. Interestingly, Bers \textit{et al.} \cite{Bers1984} treated the absolute/convective instabilities of electron beams in 3 spatial dimensions (i.e., in 4-dimensional spacetime,) relativistically in 1984 with the same method, seemingly an impossible task. Indeed they fell short of full application of the theory: the authors found a critical point but did not show that the integration contour is really pinched by the merging roots of the DR. 

This is the core of the problem. Finding critical points is an easy part of the Briggs' theory. What is most difficult is to show that a particular critical point satisfies the pinch criterion and contributes to the asymptotic behavior of the perturbation. This is because it requires knowledge on the behavior of the DR not only in the vicinity of the critical point but also in much wider a region. In the case of one spatial dimension, some people change the order of integrations and conduct the $\omega$-integral first and then apply the formula of the steepest descent method to the remaining $k$-integral rather blindly, i.e., not showing that the integral contour can be modified so that it could coincide with the steepest-decent path from the particular critical point \cite{Lingwood1997}. A disguise notwithstanding, it is apparent that this alternative formalism has actually the same problem as the partial application of Briggs' theory does. The crucial fact is that the critical point does not always contribute to the asymptotic behavior.

In this paper, we present a new formalism that circumvents the difficulty just mentioned and is applicable to arbitrary dimensions and to complicated DR's by utilizing the Lefschetz thimble method, which is based on algebraic topology and geometry and gives a generalization of the steepest descent method to complex manifolds \cite{Pham2011}. It was Witten who advocated the Lefschetz thimble method in the analysis of Chern-Simons theory \cite{Witten2010}. The method has also been applied to the sign problem in quantum chromodynamics \cite{PhysRevD.86.074506}. Our new method, also being based on the Lefschetz thimble method, can not only tell which critical points contribute to the asymptotic behavior of perturbation unambiguously but also be applied easily even when $\omega(\vec{k})$ is multi-valued, since we no longer need to solve DR $\Delta(\omega,\vec{k})=0$ as $\omega = \omega(\vec{k})$.

\section{Formulation}

\subsection{Formulation}
We consider the following system of linear partial differential equations $\mat{D}(i\partial)\vec{S}(x)=\vec{0}$ in $(d+1)$-dimensional spacetime, where $(x^\mu)\equiv(t,\vec{x})$ are the coordinates and the corresponding derivative operators with upper indices are denoted by $(\partial^\mu) \equiv \left(\eta^{\mu\nu}\frac{\partial}{\partial x^\nu}\right) = \left(\frac{\partial}{\partial t},-\frac{\partial}{\partial \vec{x}}\right)$ with the Minkowski metric $\eta \equiv \diag (1,-1,\cdots,-1)$; $\mat{D}$ is an $N\times N$ matrix with each entry being a function of the derivative operators; $\vec{S}$ is a variable to be solved and have $N$ components in general. It is typically a linear perturbation to some physical quantities of interest. We assume that the background is uniform and the elements of $\mat{D}$ are polynomials of the derivative operators with constant coefficients. This is not a serious limitation, since it is a common practice to apply linear analysis locally under the assumption that the typical wavelength of perturbation is much shorter than the scale height in the background. 

\begin{figure*}[htb]
    \begin{tabular}{cc}
        \begin{minipage}[c]{0.65\hsize}
            \subfigure[]{
                \includegraphics[width=0.47\linewidth]{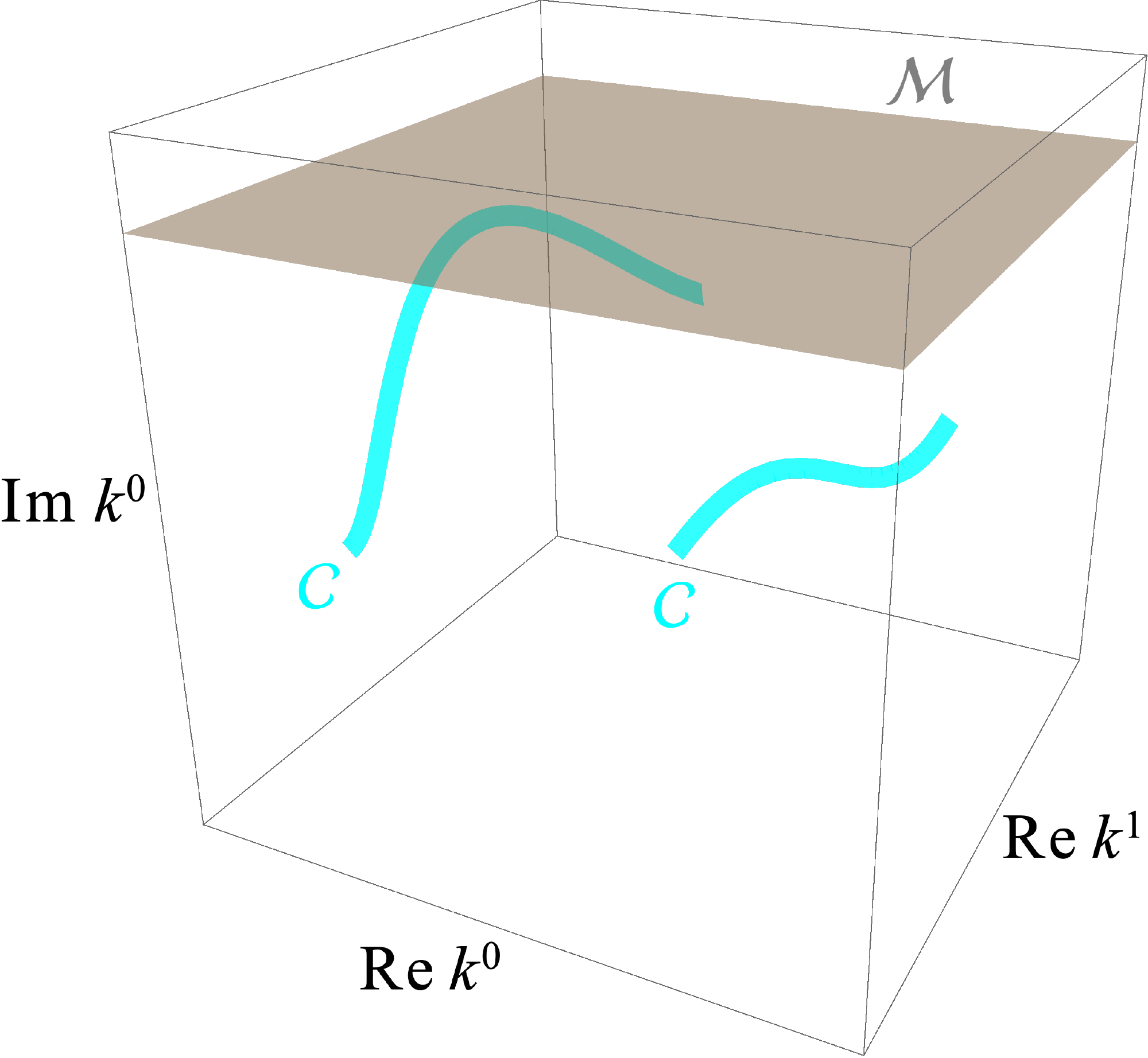}
                \label{fig:OriginalContour}
            }
            \subfigure[]{
                \includegraphics[width=0.47\linewidth]{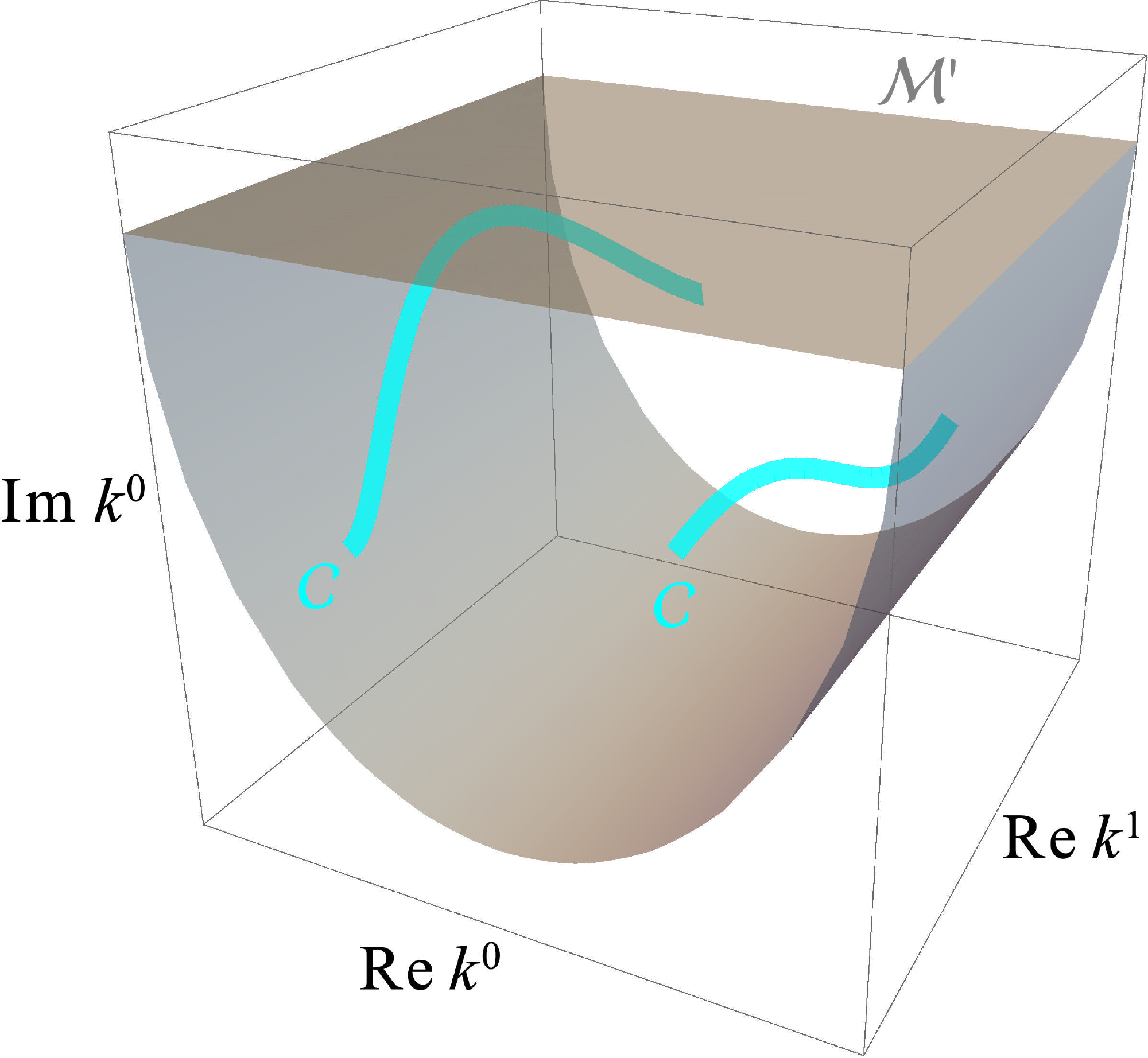}
                \label{fig:ResidueTheorem}
            }
        \end{minipage}
        \begin{minipage}[c]{0.35\hsize}
            \subfigure[]{
                \includegraphics[width=\linewidth]{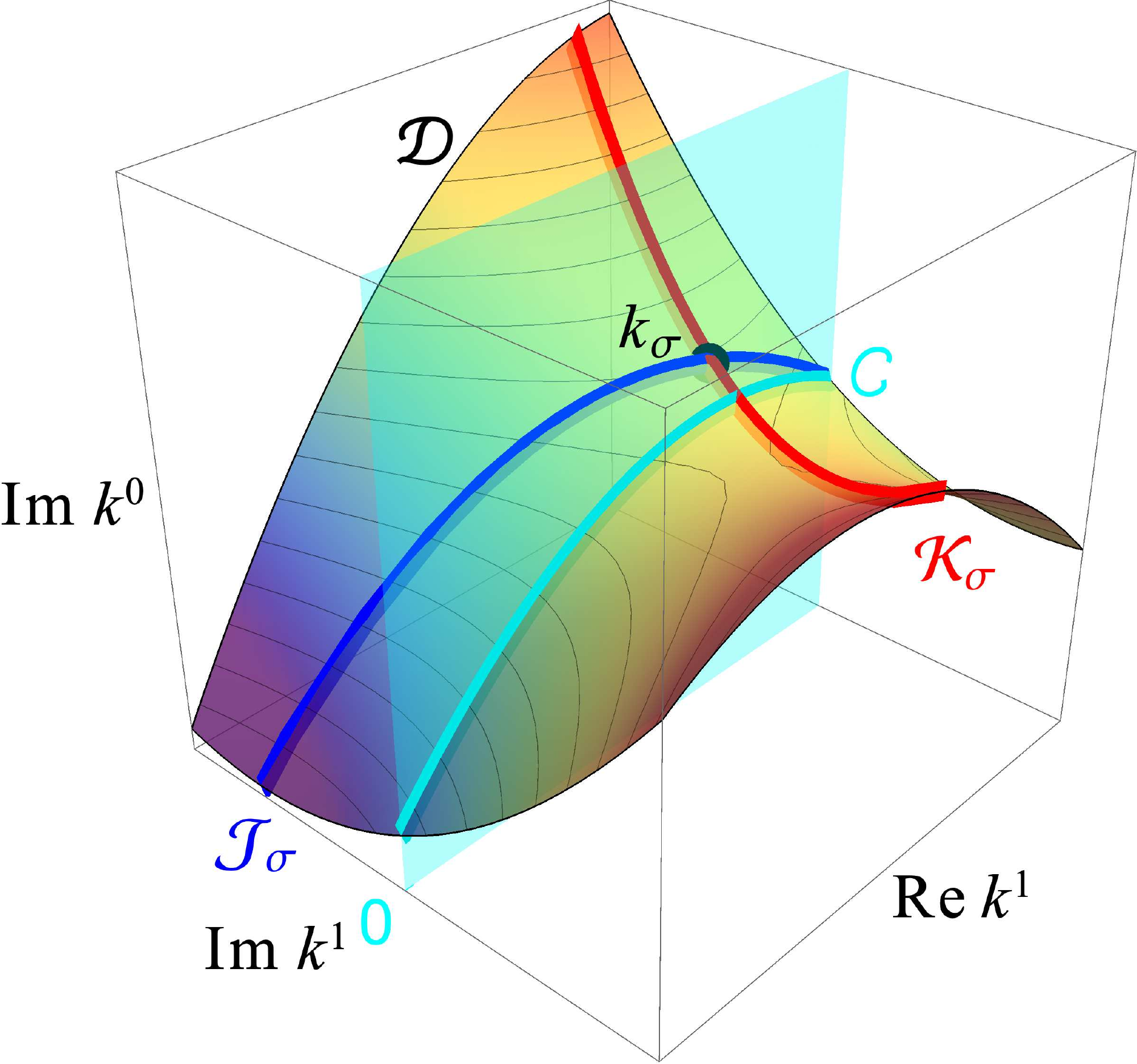}
                \label{fig:LefschetzThimbleMethod}
            }
        \end{minipage}
    \end{tabular}
    \caption{Schematic pictures of the integration contours in the complex $k$-space for the 2-dimensional spacetime. \subref{fig:OriginalContour}: $\mathcal{M}$ and $\mathcal{C}$ on the section at $\im k^1 = 0$. \subref{fig:ResidueTheorem}: Same as \subref{fig:OriginalContour} but for $\mathcal{M}'$. \subref{fig:LefschetzThimbleMethod}: Projections onto the section at $\re k^0 = 0$ of $\mathcal{C}
    $, the Lefschetz thimble $\mathcal{J}_\sigma$ and its dual thimble $\mathcal{K}_\sigma$ both attached to the critical point $k_\sigma$ on $\mathcal{D}$. The colors on $\mathcal{D}$ express the values of $h$ with reddish (purplish) hues corresponding to higher (lower) values.}
	\label{fig:SchematicPictures}
\end{figure*}

The asymptotic behavior of $\vec{S}$ is obtained by considering the retarded Green function $\mat{G}$ that satisfies $\mat{D}(i\partial)\mat{G}(x)=\delta^{(d+1)}(x)\mat{I}_N$ and the condition $\mat{G}(x)=\mat{0}$ for $t<0$. Here $\delta^{(d+1)}(x)$ is the $(d+1)$-dimensional delta function and $\mat{I}_N$ is the $N\times N$ unit matrix. The Laplace-Fourier transform enables us to express it explicitly as
\begin{align}
    \mat{G}(t,\vec{x}+\vec{v}t) = \int_{\mathcal{M}}\dfrac{d^{d+1}k}{(2\pi)^{d+1}}e^{-ik\cdot vt}e^{i\vec{k}\cdot\vec{x}}\mat{D}(k)^{-1},
    \label{eq:G}
\end{align}
where $d^{d+1}k\equiv dk^0 \wedge d^d\vec{k}$, $d^d\vec{k} \equiv dk^1 \wedge \cdots \wedge dk^d$ and $\mathcal{M}$ is the Laplace-Fourier contour (see Fig. \ref{fig:OriginalContour}), the orientation of which is given by $dk^0_r \wedge dk^1_r \wedge \cdots \wedge dk^d_r$ with $k^\mu_r \equiv \re k^\mu$; the inner product of two $(d+1)$-dimensional vectors $a$ and $b$ is denoted by $a\cdot b \equiv \eta_{\mu\nu}a^\mu b^\nu$. In the above expression the asymptotic behavior of $\mat{G}$ at $t \to 0$ is considered in a frame moving at a velocity $\vec{v}$ so that we could investigate the convective instability. Note that the absolute instability corresponds simply to the case with $\vec{v} = 0$. Incidentally $v^\mu$ is the $(d+1)$-dimensional velocity given by $v^\mu = (1, \vec{v})$. The inverse matrix of $\mat{D}(k)$ is given as $\mat{D}^{-1}(k) = \adj\mat{D}(k) / \Delta(k)$ in terms of the adjugate matrix $\adj\mat{D}(k)$ and the determinant $\Delta(k)$ of $\mat{D}(k)$ and hence has pole singularities at the zeros of $\Delta(k)$. Note that the real solutions of $\Delta(k)=0$ give the DR of stably propagating modes.

For $t>0$, the integration contour $\mathcal{M}$ can be modified to $\mathcal{M}'$ on the section at $\im k^1 = \cdots = \im k^d = 0$ by the Jordan's lemma (see Fig. \ref{fig:ResidueTheorem}) so that all the poles should be enclosed. Then the integration of Eq. (\ref{eq:G}) can be reduced to the integral of the Poincar\'e residue of the integrand by the residue theorem \cite{Griffiths1978,Pham2011}:
\begin{align}
    \mat{G}(t,\vec{x}+\vec{v}t) = \dfrac{\theta(t)}{(2\pi)^d i}\int_{\mathcal{C}}d^d\vec{k}\dfrac{e^{-ik\cdot vt}e^{i\vec{k}\cdot\vec{x}}}{\partial_0\Delta(k)}\adj\mat{D}(k),
    \label{eq:GintC}
\end{align}
where the integration contour $\mathcal{C}\equiv\mathcal{D}\cap\left(\mathbb{C}\times\mathbb{R}^d\right)$ is the section of $\mathcal{D}$ at $\im k^1 = \cdots = \im k^d = 0$, with $\mathcal{D}$ being a locus defined by $\mathcal{D}\equiv\left\{k\in\mathbb{C}^{d+1}|\Delta(k)=0\right\}$; the orientation of $\mathcal{C}$ is given by $\left.dk^1_r \wedge \cdots \wedge dk^d_r\right|_{\mathcal{C}}$; here and henceforth $\partial_\mu$ are $k$-derivatives. We note that the integrand is actually its restriction on $\mathcal{D}$ but is abbreviated for notational simplicity. The integrand is holomorphic unless there exists some $k\in \mathcal{D}$ such that $d\Delta=0$ on that point. In the following we assume $d\Delta \neq 0$ for all $k\in\mathcal{D}$, since it is satisfied indeed for almost all $\Delta$ and even if not, we could always modify $\Delta$ slightly by, for instance, adding $\epsilon k^0$ with a small $\epsilon\in\mathbb{R}$ so that it should satisfy the condition.

To obtain the asymptotic behavior of Eq. (\ref{eq:GintC}) in the limit of $t\to\infty$, we utilize the Lefschetz thimble method on $\mathcal{D}$ embedded into the $(d+1)$-dimensional complex space $\left(\mathbb{C}^{d+1},g\right)$ (see Fig. \ref{fig:LefschetzThimbleMethod}), with a K\"ahler metric $g$ satisfying $g_{\alpha\beta}=g_{\bar\alpha\bar\beta}=0,\ g_{\alpha\bar\beta}=g_{\bar\beta\alpha}=\delta_{\alpha\beta}/2$. It expresses the integration contour $\mathcal{C}$ as the sum of Lefschetz thimbles $\{\mathcal{J}_\sigma\}$, which are nothing but the steepest descent paths: $\mathcal{C} \cong \sum_\sigma\braket{\mathcal{C},\mathcal{K}_\sigma}\mathcal{J}_\sigma$, where the coefficient $\braket{\cdot,\cdot}$ is referred to as the intersection form on $\mathcal{D}$. Then Eq. (\ref{eq:GintC}) can be rewritten as
\begin{align}
    \mat{G}(t,\vec{x}+\vec{v}t) =& \dfrac{\theta(t)}{(2\pi)^d i}\sum_\sigma\braket{\mathcal{C},\mathcal{K}_\sigma}\nonumber\\
    &\times\int_{\mathcal{J}_\sigma}d^d\vec{k}\dfrac{e^{-ik\cdot vt}e^{i\vec{k}\cdot\vec{x}}}{\partial_0\Delta(k)}\adj\mat{D}(k),
    \label{eq:GintJ}
\end{align}
in which the sum runs over the critical points of a Morse function $h$ on $\mathcal{D}$ defined as the height function: $h(k) \equiv \im (k\cdot v)$, which corresponds to the real part of the exponent of $e^{-ik\cdot vt}$. Note that the intersection form ensures that only those critical points of relevance are picked up. This is exactly what was lacking in the previous theories.

There are three steps in order to evaluate the asymptotic limit of Eq. (\ref{eq:GintJ}): (1) find the critical points $\{k_\sigma\}$ of $h$ on $\mathcal{D}$; (2) obtain the intersection form $\braket{\mathcal{C},\mathcal{K}_\sigma}$; (3) evaluate the asymptotic limit of the integral of Eq. (3) on each Lefschetz thimble. There is nothing special in the first step and all one needs to do is to solve the following equations:
\begin{align}
    \begin{cases}
        \Delta (k_{\sigma}) = 0\\
        \left.dh \wedge d\Delta \wedge d\bar\Delta \right|_{k_{\sigma}} = 0,
    \end{cases}
    \label{eq:DefOfCriticalPoint}
\end{align}
which are the stationary condition for $h$ constrained on $\mathcal{D}$. After some calculations, they are written as
\begin{align}
    \begin{cases}
        \Delta (k_\sigma) = 0\\
        \left(\partial_i-v_i\partial_0\right)\Delta(k_\sigma) = 0.
    \end{cases}
    \label{eq:CriticalPoint}
\end{align}
We note that the index of $(v_\mu) = (1,-\vec{v})$ is lowered by the Minkowski metric $\eta$. 

In the second step, we construct the dual thimbles $\mathcal{K}_\sigma$ by solving ordinary differential equations. Both the Lefschetz thimbles $\mathcal{J}_\sigma$ and their dual thimbles $\mathcal{K}_\sigma$ consist of all points on the curves $K(s)$ that satisfies the upward flow equation for $h$ constrained on $\mathcal{D}$:
\begin{align}
    \dfrac{dK^\alpha(s)}{ds} = \left.\left(g^{\alpha\tilde{\beta}}-\Pi_{\re\Delta}^{\alpha\tilde{\beta}}-\Pi_{\im\Delta}^{\alpha\tilde{\beta}}\right)\partial_{\tilde{\beta}}h\right|_{k=K(s)},
    \label{eq:DefOfUpwardFlow}
\end{align}
where 
\begin{align}
\Pi_f^{\tilde{\alpha}\tilde{\beta}}(k) \equiv \dfrac{[\partial^{\tilde\alpha} f(k)][\partial^{\tilde\beta}f(k)]}{[\partial_{\tilde\gamma}f(k)][\partial^{\tilde\gamma}f(k)]}
\end{align}
is the orthogonal projector onto the hypersurface given by $f(k)=\mathrm{const.}$; the indices with tilde stand collectively for indices with and without bar; we use the following notation for contraction: $a_{\tilde \alpha}b^{\tilde \alpha} \equiv g_{\alpha\bar\beta}a^{\alpha}\overline{b^{\beta}}+g_{\bar\alpha\beta}\overline{a^{\alpha}}b^{\beta}$. 
Equation~(\ref{eq:DefOfUpwardFlow}) can be simplified by straightforward calculations as
\begin{align}
    \dfrac{dK^\alpha(s)}{ds} = iv_\beta\left[\delta^{\beta\alpha}-\delta^{\beta\gamma}\dfrac{\partial_\gamma\Delta\overline{\partial_\delta\Delta}}{\|\partial\Delta\|^2}\delta^{\delta\alpha}\right]_{k=K(s)},
    \label{eq:UpwardFlow}
\end{align}
where $\|a\| \equiv \delta_{\alpha\beta}\overline{a^\alpha} a^\beta$, and ensures that $K(s)$ satisfies the following relations for arbitrary $\vec{v}$ and $s$:
\begin{itemize}
    \item the constraint on $\mathcal{D}$:
    \begin{align}
        \Delta(K(s)) = 0
    \end{align}
    \item the stationarity of the phase of $e^{-ik\cdot vt}$:
    \begin{align}
        \frac{d}{ds}\re (K(s)\cdot v) = 0
    \end{align}
    \item the monotonicity of the amplitude of $e^{-ik\cdot vt}$:
    \begin{align}
        \frac{d}{ds}\im (K(s)\cdot v) = \frac{d}{ds}h(K(s),\vec{v}) \geq 0.
    \end{align}
\end{itemize}
The difference between $\mathcal{J}_\sigma$ and $\mathcal{K}_\sigma$ is the boundary condition they satisfy: $K(\infty) = k_\sigma$ for $\mathcal{J}_\sigma$ and $K(-\infty) = k_\sigma$ for $\mathcal{K}_\sigma$. Note that the orientations of $\mathcal{J}_\sigma$ and $\mathcal{K}_\sigma$ are chosen so that the following orthogonal relations should hold: $\braket{\mathcal{J}_\sigma,\mathcal{K}_{\sigma'}} = \delta_{\sigma\sigma'}$. It is incidentally mentioned that since $h$ depends on $\vec{v}$, so do $k_\sigma$, $\mathcal{J}_\sigma$ and $\mathcal{K}_\sigma$. We do not show it explicitly, though, for notational simplicity unless it is important.

\begin{figure*}[htb]       
    \subfigure[]{
        \includegraphics[width=0.46\linewidth]{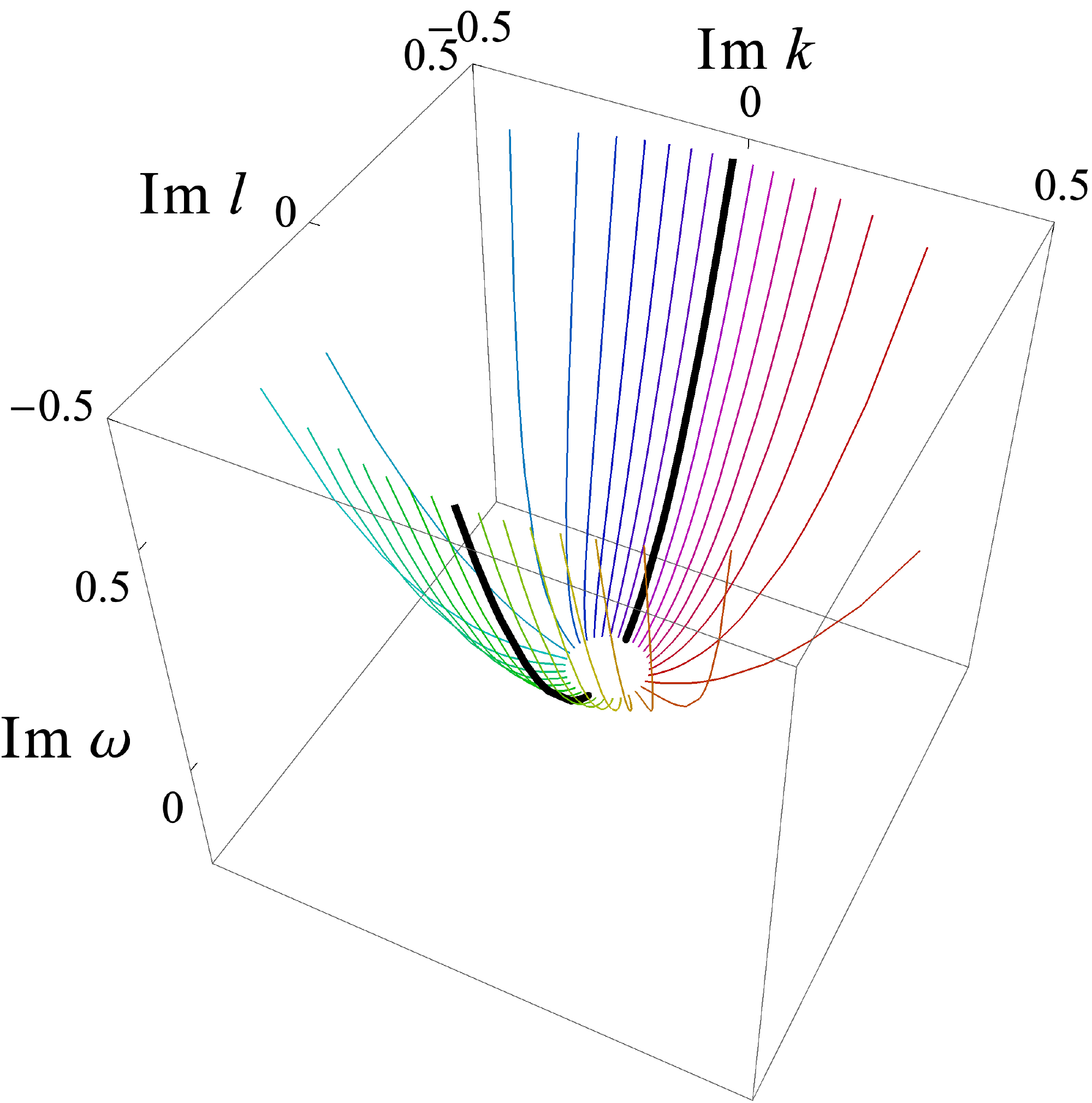}
        \label{fig:UpwardFlowsSmallScale}
    }
    \subfigure[]{
        \includegraphics[width=0.46\linewidth]{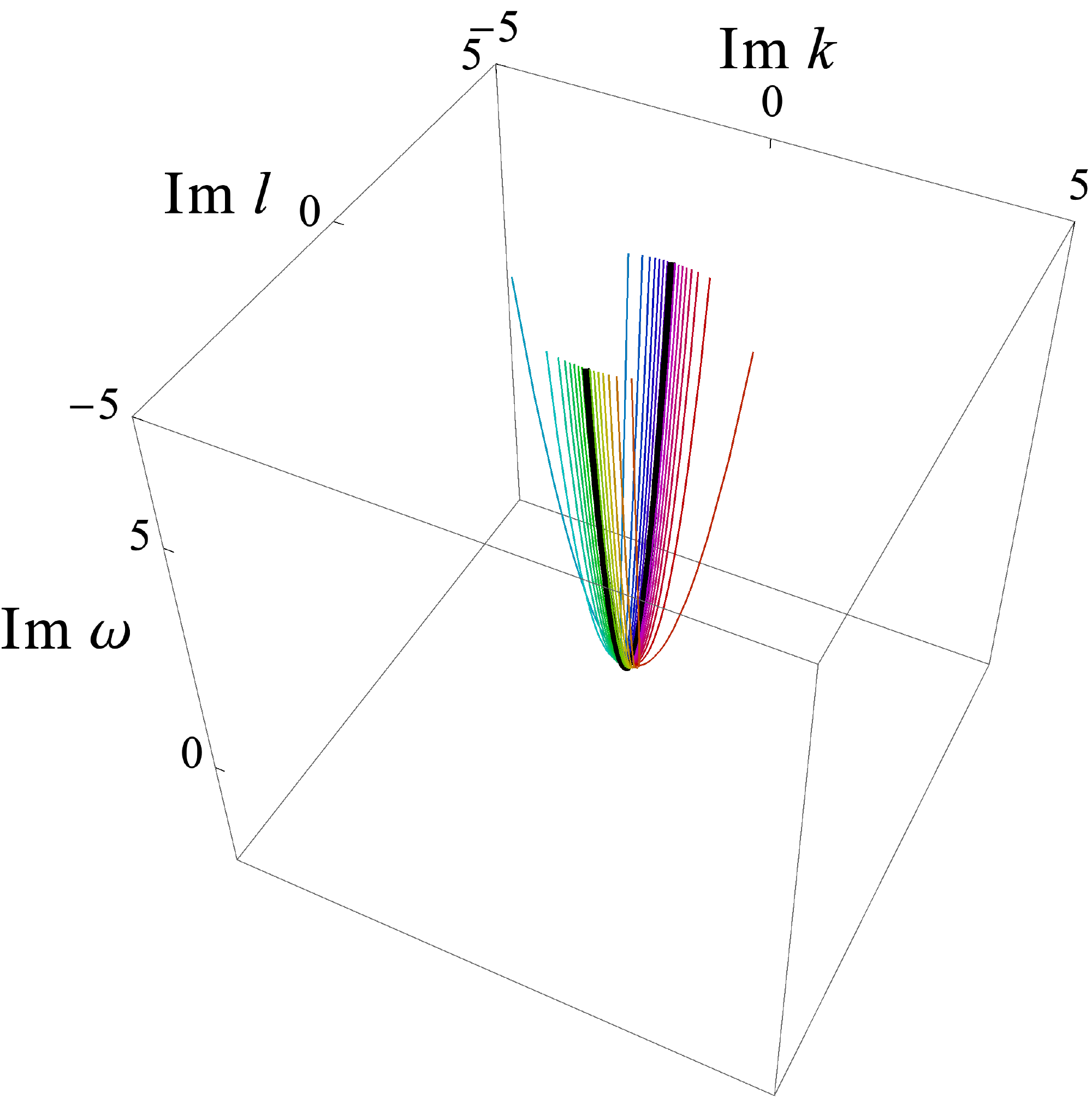}
        \label{fig:UpwardFlowsLargeScale}
    }
	\caption{Schematic pictures of upward flow lines~\footnote{In these figures, the upward flow lines are in fact drawn by solving the gradient flow equations (\ref{eq:UpwardFlow}) for the height function $h(\omega,k,l) = \im\omega$ with the constraint $\Delta(\omega,k,l) = k^2+2l^2-\omega = 0$.}. The black solid lines are the steepest flow lines. \subref{fig:UpwardFlowsLargeScale} shows the same as \subref{fig:UpwardFlowsSmallScale} in the larger scale than \subref{fig:UpwardFlowsSmallScale}.}
	\label{fig:UpwardFlows}
\end{figure*}

One needs to solve (numerically in general) the flow equation (\ref{eq:UpwardFlow}) with the initial condition $K(-\infty)=k_\sigma$ for each critical point to construct the dual thimbles attached to the critical point $k_\sigma$ and obtain the intersection form $\braket{\mathcal{C},\mathcal{K}_\sigma}$. This is the cost we have to pay in this method to avoid the costly and almost impossible exploration of the complex DR over the entire complex $k$-space. As a technical tip, we suggest to set the initial condition for the integration as $K(0) = k_\sigma + \kappa$, where a small shift $\kappa$ is chosen as
\begin{align}
    \begin{cases}
        \vec{\kappa} \equiv i\mat{J}_\sigma \vec{\delta}\\
        \kappa^0 \equiv \vec{v}\cdot\vec{\kappa},
    \end{cases}
    \label{eq:Perturbation}
\end{align}
where $\mat{J}_\sigma$ is given below and $\vec{\delta}\in\mathbb{R}^3$ is a small real vector. This prescription ensures that $K(0)$ sits on the dual thimble $\mathcal{K}_\sigma$. For each $\vec{\delta}$ one obtains a single upward flow line and $\mathcal{K}_\sigma$ is the collection of all these upward flow lines. We do not need all of them in fact.
It should be noted that many flow lines seem to be attracted to the steepest flow lines as $s$ evolves as shown in Fig.~\ref{fig:UpwardFlows}.
Then the whole picture of a dual thimble may not be captured unless the upward flow lines are sufficiently many. 
This concentration of flow lines occurs at smaller scales for smaller perturbations of $\kappa$ indeed.
We should hence choose appropriately large (and not too large of course) perturbations;
otherwise large numbers of flow lines are required to recover the entire shape of a dual thimble.

The intersection form $\braket{\mathcal{C},\mathcal{K}_\sigma}$ is easily evaluated as follows. We focus on the imaginary parts of the spatial coordinates of $\mathcal{C}$ and $\mathcal{K}_\sigma$, i.e., we project them onto the section at $k^0 = \re k^1 = \cdots = \re k^d = 0$. Then $\mathcal{C}$ is reduced to the origin $\im k^1 = \cdots = \im k^d = 0$ whereas the projection of $\mathcal{K}_\sigma$ is still $d$-dimensional in general. The evaluation of $\braket{\mathcal{C},\mathcal{K}_\sigma}$ is hence reduced to the investigation of whether the projection of $\mathcal{K}_\sigma$ includes the origin or not. 
In other words, $\braket{\mathcal{C},\mathcal{K}_\sigma} = \pm 1$ if the section of $\mathcal{K}_\sigma$ at $s\to\infty$ (or sufficiently large $s$ practically), which is isomorphic to $S^{d-1}$, encloses the origin in the projected space; 
$\braket{\mathcal{C},\mathcal{K}_\sigma} = 0$ otherwise.
Importantly from a practical point of view, this can be visualized for the spacetime dimension less than 5.

The last step is to evaluate the integral on each Lefschetz thimble in the limit of $t\to\infty$. Since the Lefschetz thimble is nothing but a steepest descent path from a critical point, the integral is dominated by the contribution from the vicinity of the critical point. Then the asymptotic form of $\mat{G}$ for $|k_\sigma \cdot v t| \gg |\vec{k}_\sigma\cdot \vec{x}|$ is given as
\begin{align}
    \mat{G}(t,\vec{x}+\vec{v}t) \sim& \dfrac{1}{(2\pi t)^{d/2} i} \sum_\sigma \braket{\mathcal{C},\mathcal{K}_\sigma}e^{-ik_\sigma \cdot vt}\nonumber\\
    &\times \dfrac{e^{i\vec{k}_\sigma\cdot \vec{x}}\det\mat{J}_\sigma}{\partial_0\Delta(k_\sigma)}\adj \mat{D}(k_\sigma).
    \label{eq:AsymptoticLimitOfG}
\end{align}
In deriving this expression, we expand the exponent of $e^{-ik\cdot vt}$ up to the quadratic order at the critical point and conduct the resultant Gaussian integrals, assuming that all  other factors are subdominant and ignoring their variations; the matrix $\mat{J}_\sigma = (\vec{\iota}_1,\cdots,\vec{\iota}_d)$ is defined so that it should satisfy $\mat{H}_\sigma = -^t\mat{J}_\sigma^{-1}\mat{J}_\sigma^{-1}$ for the following matrix
\begin{align}
    \left(H_\sigma\right)_{ij}\equiv \left.i\dfrac{(\partial_i-v_i\partial_0)(\partial_j-v_j\partial_0)\Delta}{\partial_0 \Delta}\right|_{k=k_\sigma}
\end{align}
and the order of the column vectors $\{\vec{\iota}_i\}$ are chosen so that $\iota_1 \wedge \cdots \wedge \iota_d$ should give the orientation of the Lefschetz thimble at $k_\sigma$, where we introduce the $(d+1)$-dimensional vectors $\iota_i \equiv (\vec{v}\cdot \vec{\iota}_i,\vec{\iota}_i)$ associated with $\vec{\iota}_i$ so that they should be tangential to $\mathcal{D}$. Note also that the Hessian of $h$ at $k_\sigma$ on $\mathcal{D}$ is then given by $\re\left[\left(H_\sigma\right)_{ij}dk^i\otimes dk^j\right]$ (see \cite{Birtea2015} for the derivation). 
The choice of $\mat{J}_\sigma$ is not unique and the results are unchanged by the transformation $\mat{J}_\sigma \mapsto \mat{J}_\sigma \mat{V}$ for $\mat{V}\in \mathrm{SO}(d,\mathbb{C})$. Moreover, since one can express $\det\mat{J}_\sigma$ as $\det \mat{J}_\sigma = e^{i\arg\det \mat{J}_\sigma}\left|\det \mat{H}_\sigma\right|^{-1/2}$ and its argument is invariant under the transformation $\mat{J}_\sigma \mapsto \mat{J}_\sigma \mat{M}$ for $\mat{M}\in\mathrm{GL}(d,\mathbb{R})$, the absolute value and argument of $\det\mat{J}_\sigma$ are obtained, respectively, from $|\det\mat{H}_\sigma|$ and conveniently chosen $\mat{J}_\sigma$ that is still amenable to the orientation of $\mathcal{J}_\sigma$.

Finally, if $\det\mat{H}_\sigma = 0$ at some $k_\sigma$, it means that $h$ is not a Morse function and the formulation given so far cannot be applied at this point. Even in this case, however, $h$ can be modified to become a Morse function just by changing $\vec{v}$ infinitesimally, since Morse functions are densely existent. Hence our results are always valid essentially.

\subsection{Maximum growth rate}
The asymptotic limit of $\mat{G}(t,\vec{x}+\vec{v}t)$ for $t\to\infty$, Eq. (\ref{eq:AsymptoticLimitOfG}), is valid for arbitrary $\vec{v}$. Then $h(k_\sigma,\vec{v})=\im(k_\sigma\cdot v)$ can be regarded as the growth rate of the contribution from a critical point $k_\sigma$. The natural question next should be in which frame the growth rate of instability becomes maximum. The answer can be obtained simply as follows if the $\im k^0$ has a maximum in $\mathcal{C}$. 

We first define $k_{\mathrm{m}}$ as 
\begin{align}
    k_{\mathrm{m}} \equiv \argsup_{k\in\mathcal{C}}\im k^0 = \argsup_{k\in\mathcal{C}}h(k,\vec{v})\ \ (^\forall\vec{v}\in\mathbb{R}^d),
\end{align}
where the last equality is satisfied because $h(k, \vec{v}) = \im(k\cdot v) = \im(k^0 - \vec{k} \cdot \vec{v})$ and  $\vec{k}\in \mathbb{R}^d$ for $k\in\mathcal{C}$. It is actually one of the critical points for the frame moving at the velocity is
\begin{align}
    \vec{v}_{\mathrm{m}} \equiv \left(-\left.\dfrac{\partial_i\Delta}{\partial_0\Delta}\right|_{k_{\mathrm{m}}}\right) \in \mathbb{R}^d,
\end{align}
since Eq. (\ref{eq:CriticalPoint}) is satisfied at $k_{\mathrm{m}}$ indeed. In this frame, the dual thimble $\mathcal{K}_{\mathrm{m}}(\vec{v}_{\mathrm{m}})$ associated with this critical point intersects with $\mathcal{C}$ at $k=k_{\mathrm{m}}$ as the critical point is sitting on $\mathcal{C}$. We hence have $\braket{\mathcal{C},\mathcal{K}_{\mathrm{m}}(\vec{v}_{\mathrm{m}})}=\pm 1$. The corresponding growth rate is $h(k_{\mathrm{m}},\vec{v}_{\mathrm{m}}) = \im k^0_{\mathrm{m}}$ by definition. 

Now we see that $\mathcal{C}$ has no intersection with any dual thimble $\mathcal{K_\sigma}(\vec{v})$ associated with those critical points $k_\sigma(\vec{v})$ that satisfy
\begin{align}
    h(k_\sigma,\vec{v}) > \im k_{\mathrm{m}}^0 = \sup_{k\in\mathcal{C}}h(k,\vec{v})\ \ (\vec{v}\neq \vec{v}_{\mathrm{m}}).
\end{align}
This is because, if otherwise, the above inequality contradicts the relation $h(k,\vec{v}) \geq h(k_\sigma,\vec{v})$, which should hold for all points on $\mathcal{K}_\sigma(\vec{v})$. Therefore, $h(k_\sigma,\vec{v}) \leq \im k_{\mathrm{m}}^0$ is satisfied for all critical points that contribute to the sum in Eq. (\ref{eq:AsymptoticLimitOfG}), which implies that the growth rate takes its maximum value $\im k_{\mathrm{m}}^0$ in the frame with the velocity $\vec{v}_{\mathrm{m}}$.

Although the maximum growth of instability is certainly one of the most relevant quantities, we stress that we are equally interested in the entire picture of the instabilities, convective and absolute alike, i.e., we want to know in which frame the instability grows at what rate. Such complete information can be also obtained in our formulation by evaluating Eq. (\ref{eq:AsymptoticLimitOfG}) for all critical points.

\section{Demonstration}

\subsection{Spatially 2-dimensional dispersion relation}
\begin{figure}[htb]       
    \includegraphics[width=0.92\linewidth]{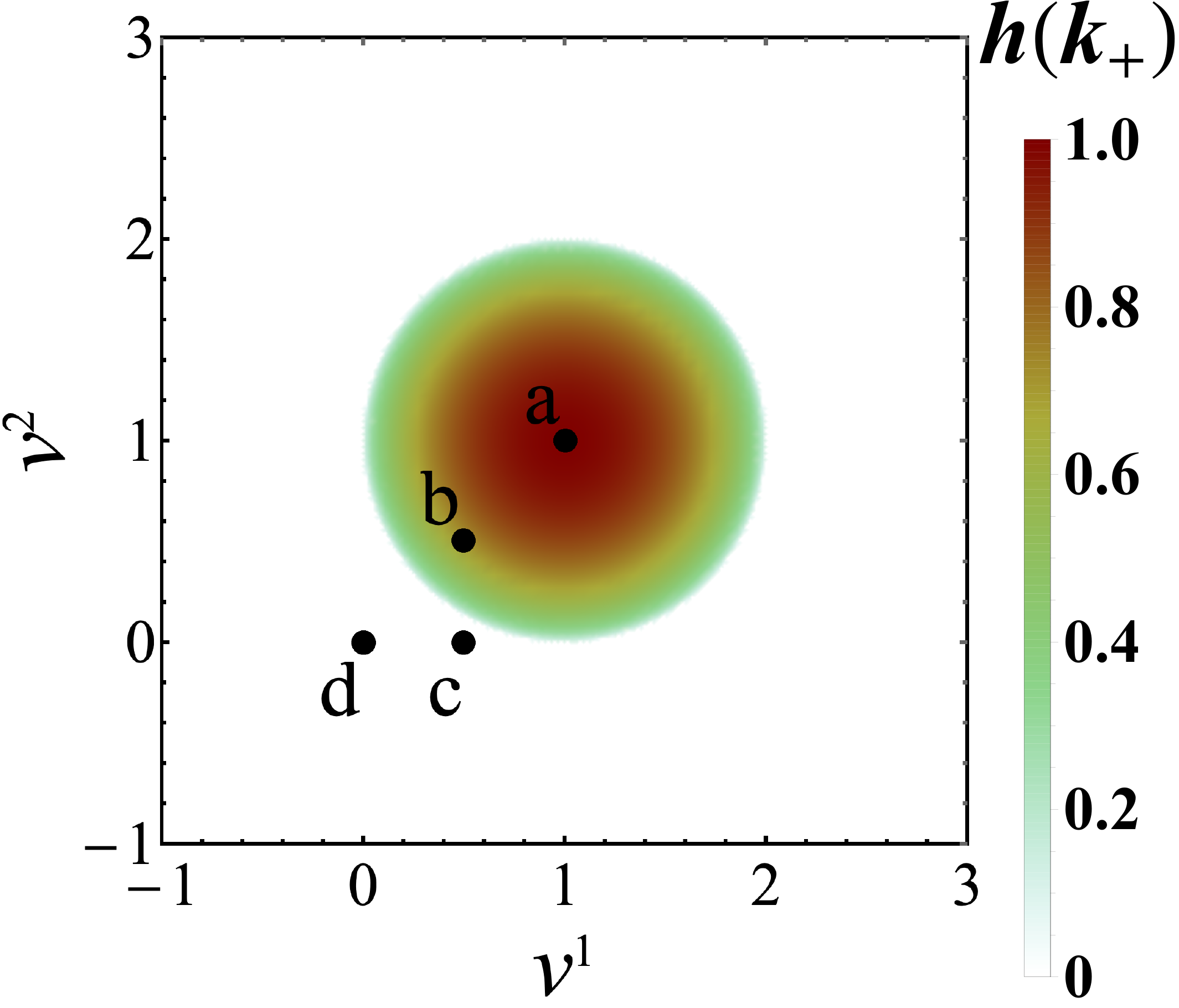}
    \caption{Color contour plot in the $v^1-v^2$ plane of the Morse function $h(k,\vec{v})$ at $k_+$.}
    \label{fig:GrowthRateMap}
\end{figure}

Here we demonstrate how our new method works, employing the following DR given in \cite{Brevdo1991}:
\begin{align}
    \Delta(k) = \left(k^1\right)^2 + \left(k^2\right)^2 - \left(k^0 - k^1 - k^2\right)^2 + 1 = 0.
    \label{eq:ToyDR}
\end{align}
This toy model has a merit that it can be treated completely analytically. In fact the Morse function $h(k,\vec{v})$ is shown in Fig. \ref{fig:GrowthRateMap} as a function of $\vec{v}$ at the critical point $k_+$, which gives the highest value to $h$. It is found that $h(k_+,\vec{v})$ is positive only when $(v^1-1)^2+(v^2-1)^2<1$, which is hence a necessary condition for instability for this DR.

\begin{figure*}[htb]       
    \subfigure[$\vec{v}=(1,1)$]{
        \includegraphics[width=0.46\linewidth]{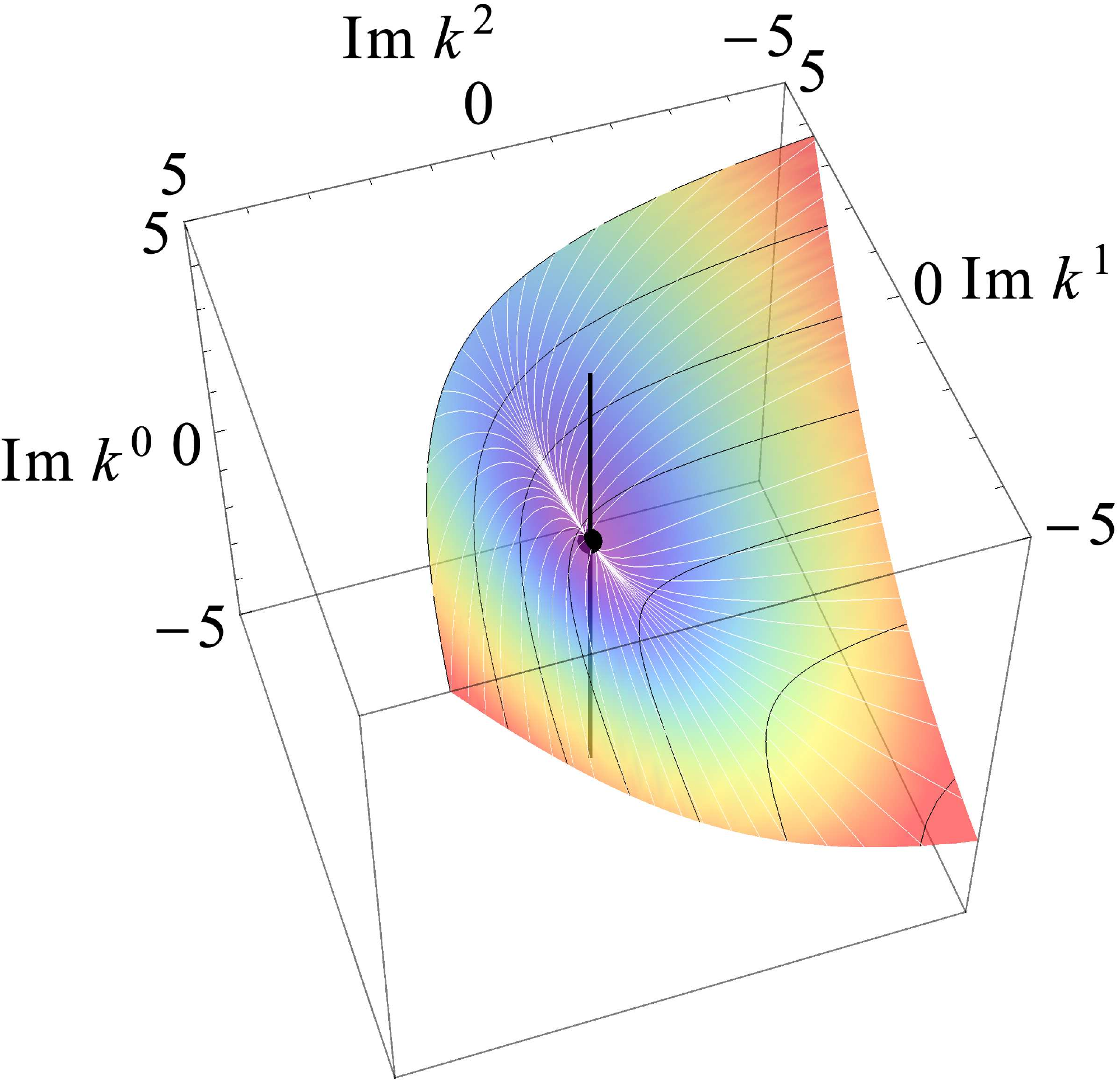}
        \label{fig:DualThimble11}
    }
    \subfigure[$\vec{v}=(0.5,0.5)$]{
        \includegraphics[width=0.46\linewidth]{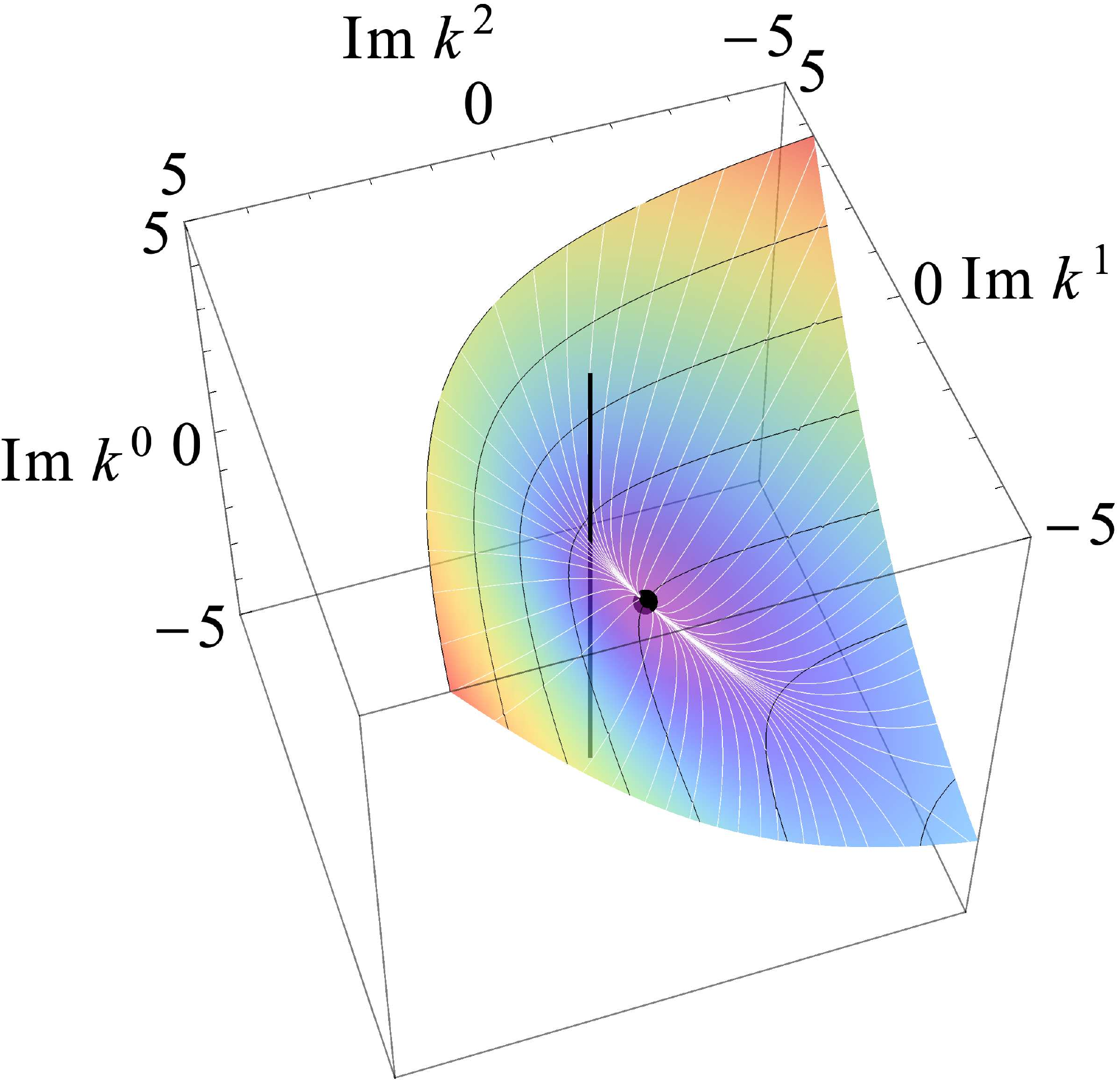}
        \label{fig:DualThimble0505}
    }
    \subfigure[$\vec{v}=(0.5,0)$]{
        \includegraphics[width=0.46\linewidth]{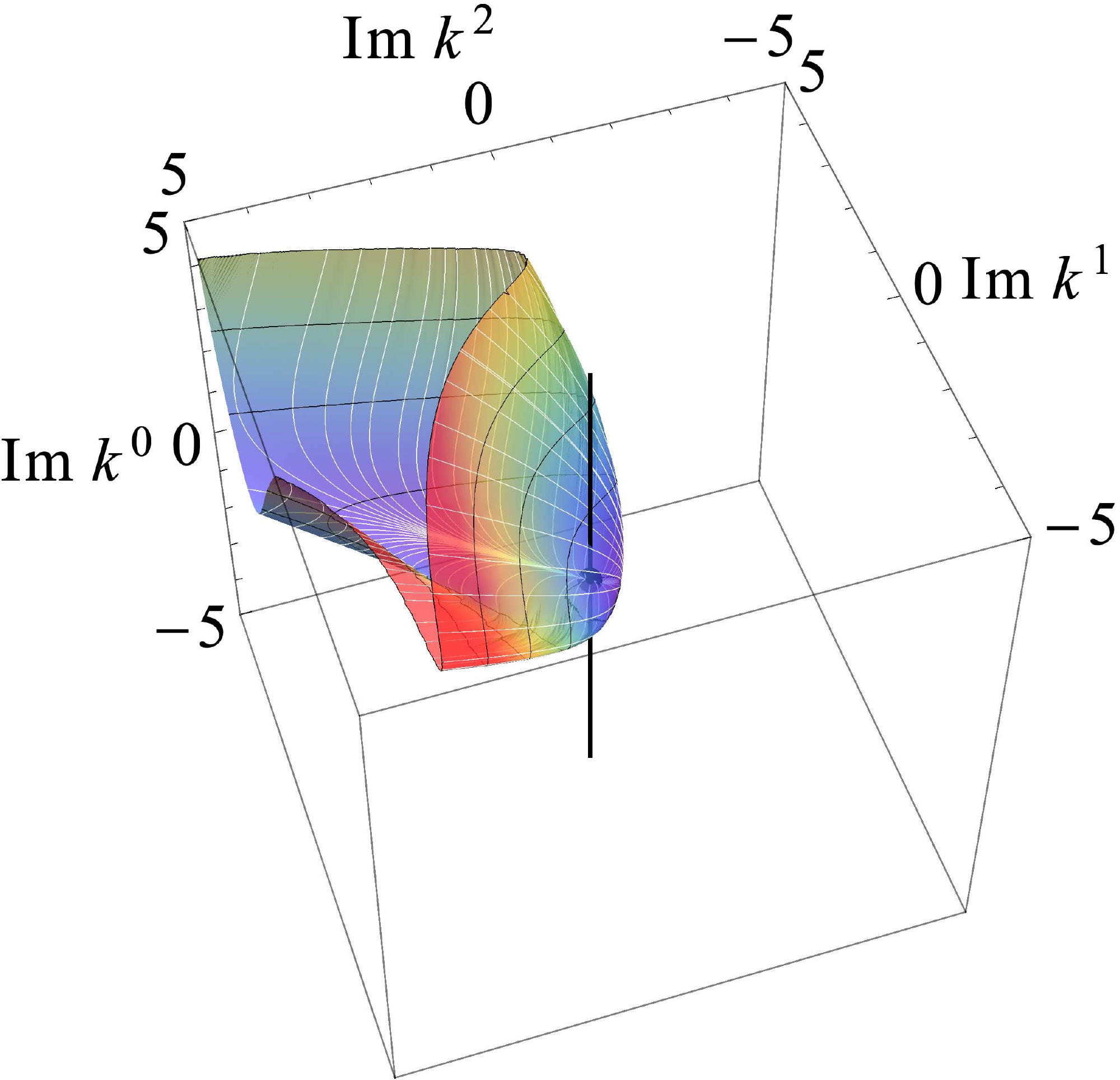}
        \label{fig:DualThimble050}
    }
    \subfigure[$\vec{v}=(0,0)$]{
        \includegraphics[width=0.46\linewidth]{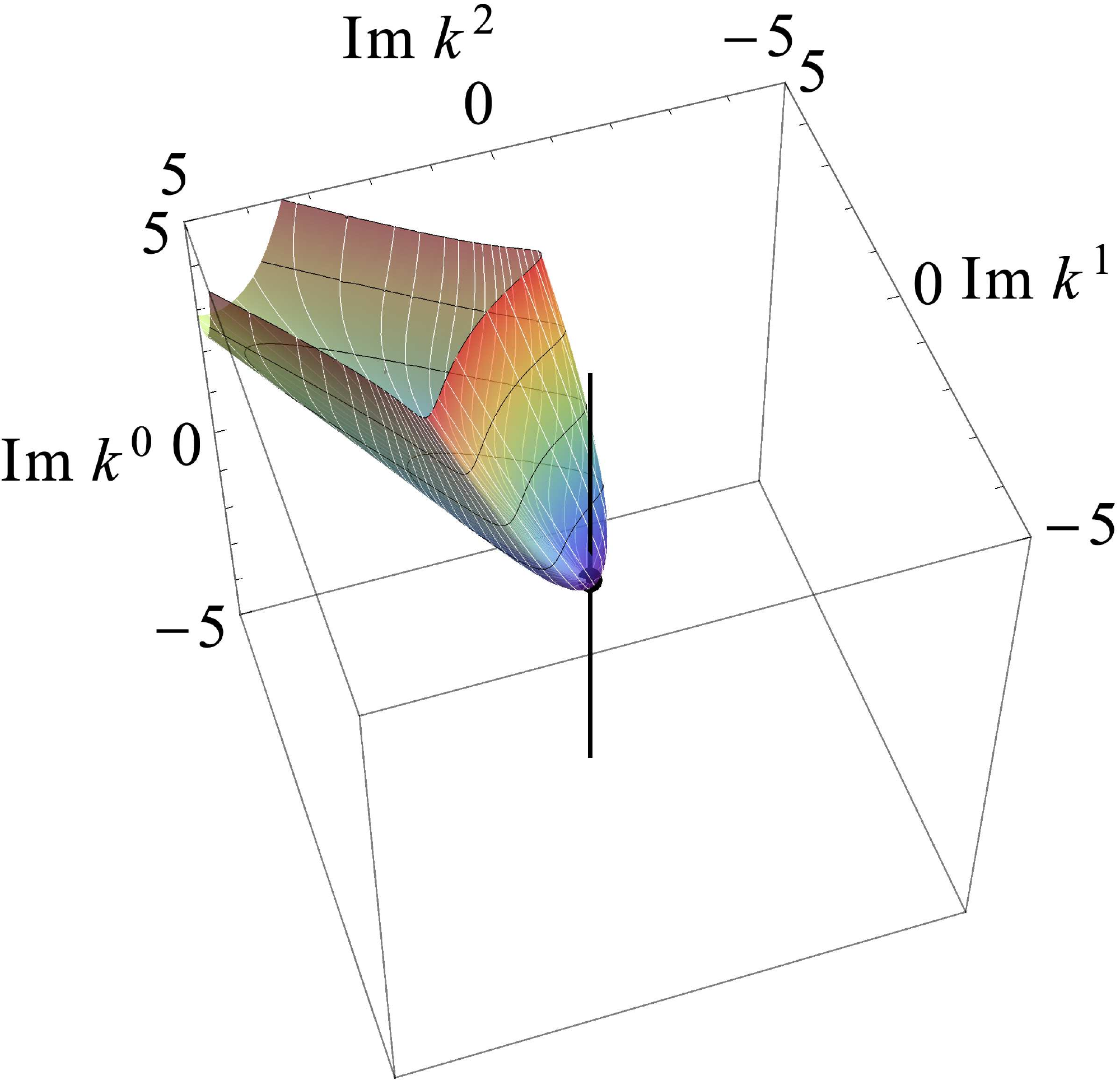}
        \label{fig:DualThimble00}
    }
    \caption{The dual thimbles $\mathcal{K}_+$ attached to the critical points $k_+$ for the DR given by Eq.~(\ref{eq:ToyDR}) for the values of $\vec{v}$ specified in Fig.~\ref{fig:GrowthRateMap}. The black dots are $k_+$, the vertical black lines indicates the points with $\im k^1 = \im k^2 =0$, parts of which correspond to $\mathcal{C}$ and the rainbow surfaces are $\mathcal{K}_+$. The purplish (reddish) colors stand for low (high) values of $h$ and white lines are representative flow lines. }
	\label{fig:DualThimbles}
\end{figure*}

To study whether this is also the sufficient condition or not in the classical approach, we have to accomplish the following tasks for each $\vec{v}$ in the range given above. We first solve the simultaneous equations $\{\Delta(k)=0,\ \partial_1\Delta(k)=0\}$ for $k^0=\{k^0_\sigma+is|s\geq 0\}$ with an arbitrary positive $s$ for every critical point so that the trajectories $(k^1(k^0),k^2(k^0))$ could be obtained. We then investigate for all points $k=\tilde{k}$ on the trajectories whether there occurs a coalescence between two of the roots $k^1(k^0,\tilde{k}^2)$ of $\Delta(k^0,k^1,\tilde{k}^2)=0$ at $k^0 = \tilde{k}^0$. If it is true, those points on the trajectories are called coalescence roots. Finally, we have to survey further whether two of such trajectories $(k^1(k^0),k^2(k^0))$ of the coalescence roots meet at one of the critical points or not. Although all these tasks can be accomplished for the current toy model, in which the DR is fairly simple and the spatial dimension is just 2, it is not difficult to imagine how tough that would be for more complicated DR's and/or higher spatial dimensions.

In our new formulation, the procedure is simpler. 
We have only to solve Eq.~(\ref{eq:UpwardFlow}) with the initial condition $K(-\infty) = k_\sigma$ to construct the dual thimble $\mathcal{K}_\sigma$ for each critical point to find the intersection form $\braket{\mathcal{C},\mathcal{K}_\sigma}$ by examining whether $\mathcal{K}_\sigma$ includes the origin in the $\im k^1-\im k^2$ plane. 

We can then verify that the condition $(v^1-1)^2+(v^2-1)^2<1$ is indeed the sufficient condition. 
We show in Figs.~\ref{fig:DualThimble11}, \subref{fig:DualThimble0505}, \subref{fig:DualThimble050} and \subref{fig:DualThimble00} the critical point $k_+$ as well as the corresponding $\mathcal{K}_+$ obtained by solving Eq.~(\ref{eq:UpwardFlow}) numerically. 
In Figs.~\ref{fig:DualThimble11} and \subref{fig:DualThimble0505}, $\mathcal{K}_+$ intersects with $\mathcal{C}$ only once, implying $\braket{\mathcal{C},\mathcal{K}_+} = \pm 1$ and that this critical point contributes to the asymptotic behavior. 
On the other hand, in Figs.~\ref{fig:DualThimble050} and \subref{fig:DualThimble00}, the intersection of $\mathcal{K}_+$ with $\mathcal{C}$ occurs twice with opposite orientations and, as a result, $\braket{\mathcal{C},\mathcal{K}_+} = 0$ in these cases. 
We can also confirm that the dual thimble $\mathcal{K}_-$ associated with the other critical point $k_{-}$ has $\braket{\mathcal{C},\mathcal{K}_-} = 0$. 
It turns out that the Green function $\mat{G}(t,\vec{x}+\vec{v}t)$ is exactly $\mat{0}$ at $\vec{x}=\vec{0}$ for these values of velocity $\vec{v}$. 
Note that although it is plotted in Figs.~\ref{fig:DualThimble11}-\subref{fig:DualThimble00}, $\im k^0$ is not necessary in fact to obtain the intersection form $\braket{\mathcal{C},\mathcal{K}_\sigma}$.

\subsection{Spatially 3-dimensional dispersion relation}
Next we demonstrate our new method for spatially 3-dimensional problem with the DR
\begin{align}
    \Delta(k) =& \left(k^1\right)^2 + \left(k^2\right)^2 + \left(k^3\right)^2 \nonumber\\
    &- \left(k^0 - k^1 - k^2 - k^3\right)^2 + 1 = 0,
    \label{eq:ToyDR3D}
\end{align}
which is a straightforward extension of Eq.~(\ref{eq:ToyDR}) to 4-dimensional spacetime. 
The necessary condition for instability, under which $h(k_+,\vec{v})$ takes a positive value, is $(v^1-1)^2+(v^2-1)^2+(v^3-1)^2<1$ for this DR.

\begin{figure*}[htb]       
    \includegraphics[width=0.46\linewidth]{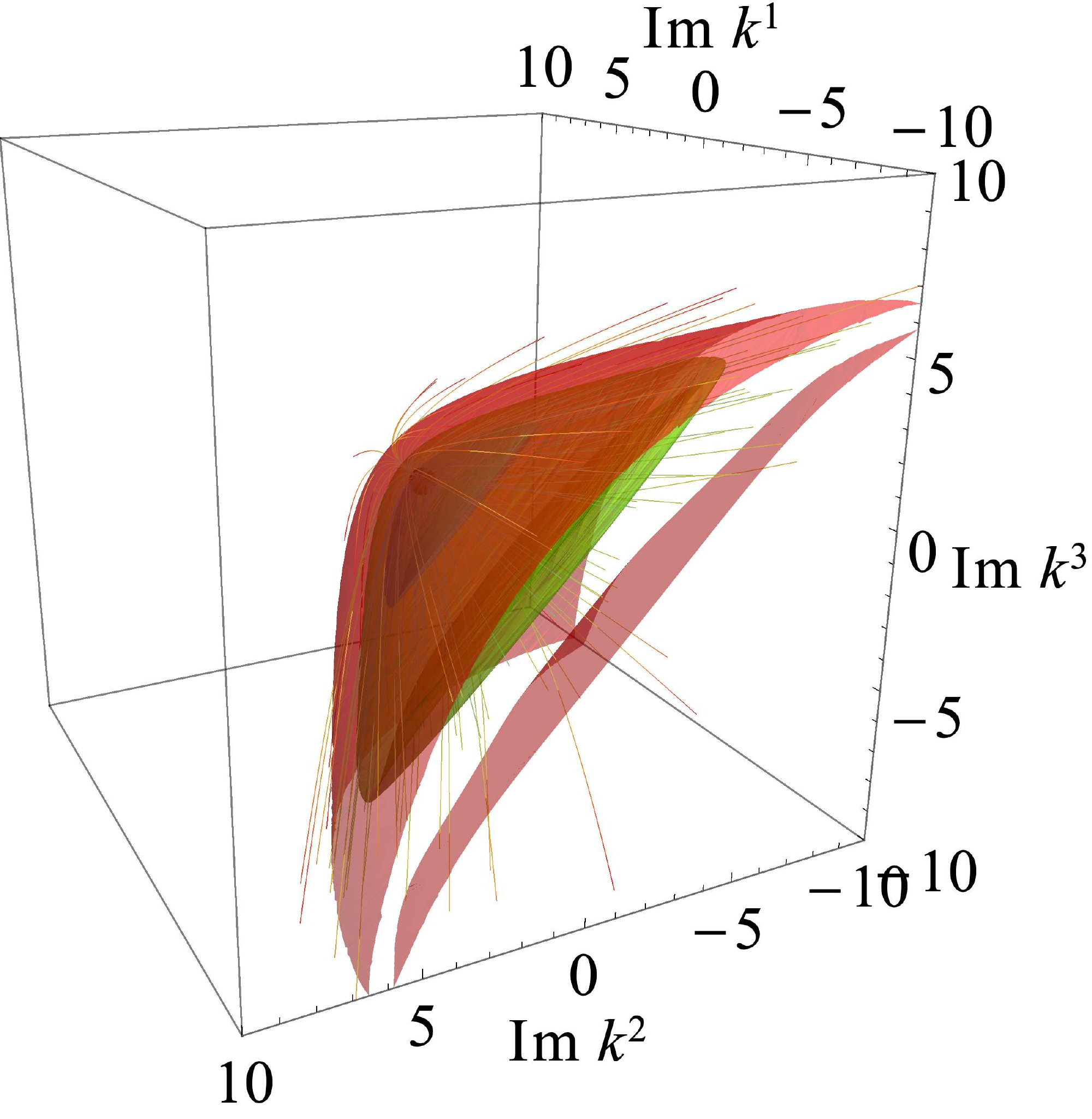}
    \includegraphics[width=0.46\linewidth]{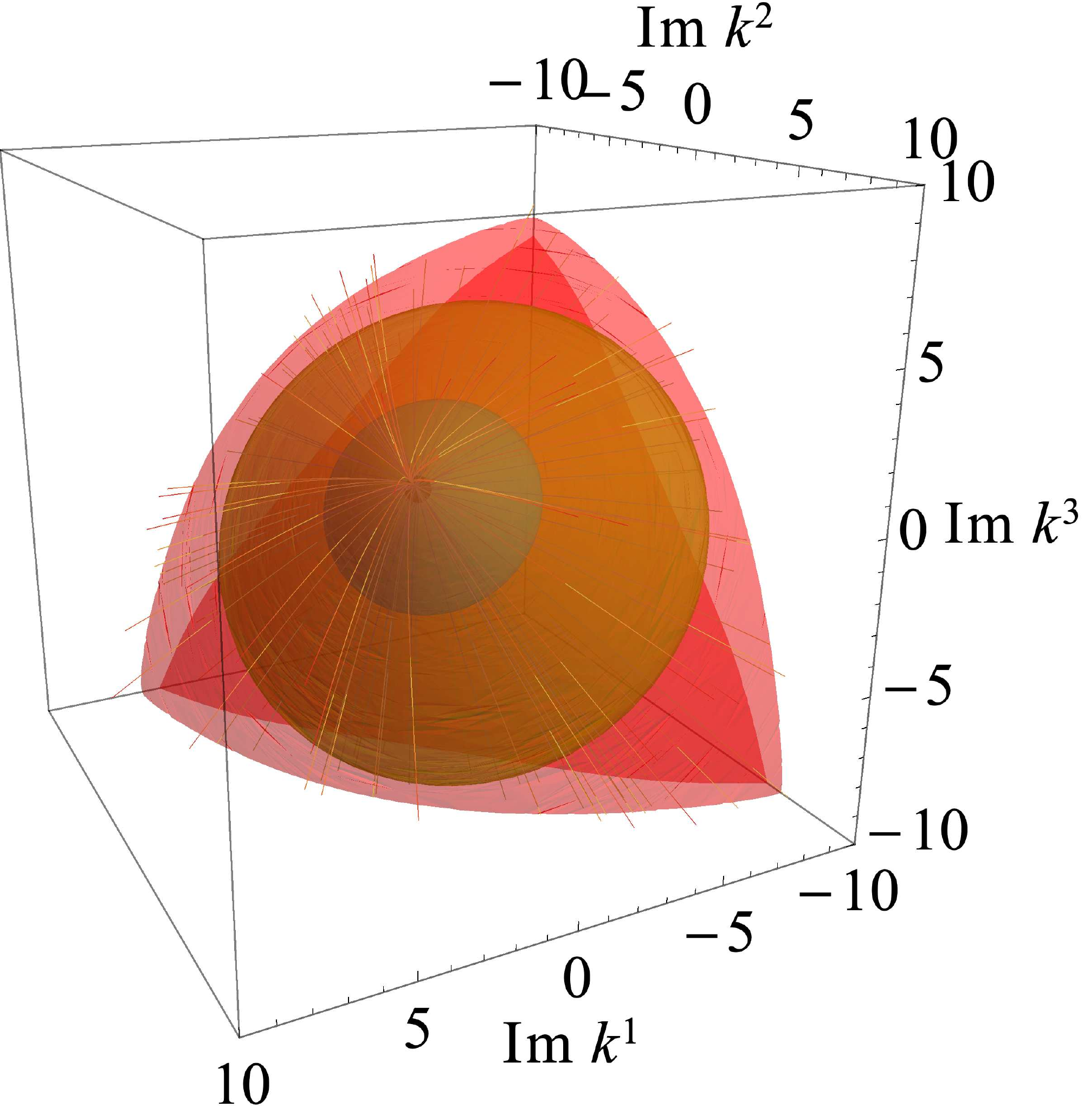}
    \caption{The dual thimbles $\mathcal{K}_+$ attached to the critical points $k_+$ for the DR given by Eq.~(\ref{eq:ToyDR3D}) for $\vec{v}=(1,1,1)$. The black dots are the origins, which indicate $\im k^1 = \im k^2 = \im k^3 = 0$. The blue, green and red surfaces are the sections of $\mathcal{K}_+$ at some $s$, where the blue (red) is for the smallest (largest) $s$. The rainbow lines are representative flow lines, where the purplish (reddish) colors stand for small (large) values of $s$. The left and right panels show the same things from the different angles.}
    \label{fig:DualThimble3D111}
\end{figure*}

\begin{figure*}[htb]       
    \includegraphics[width=0.46\linewidth]{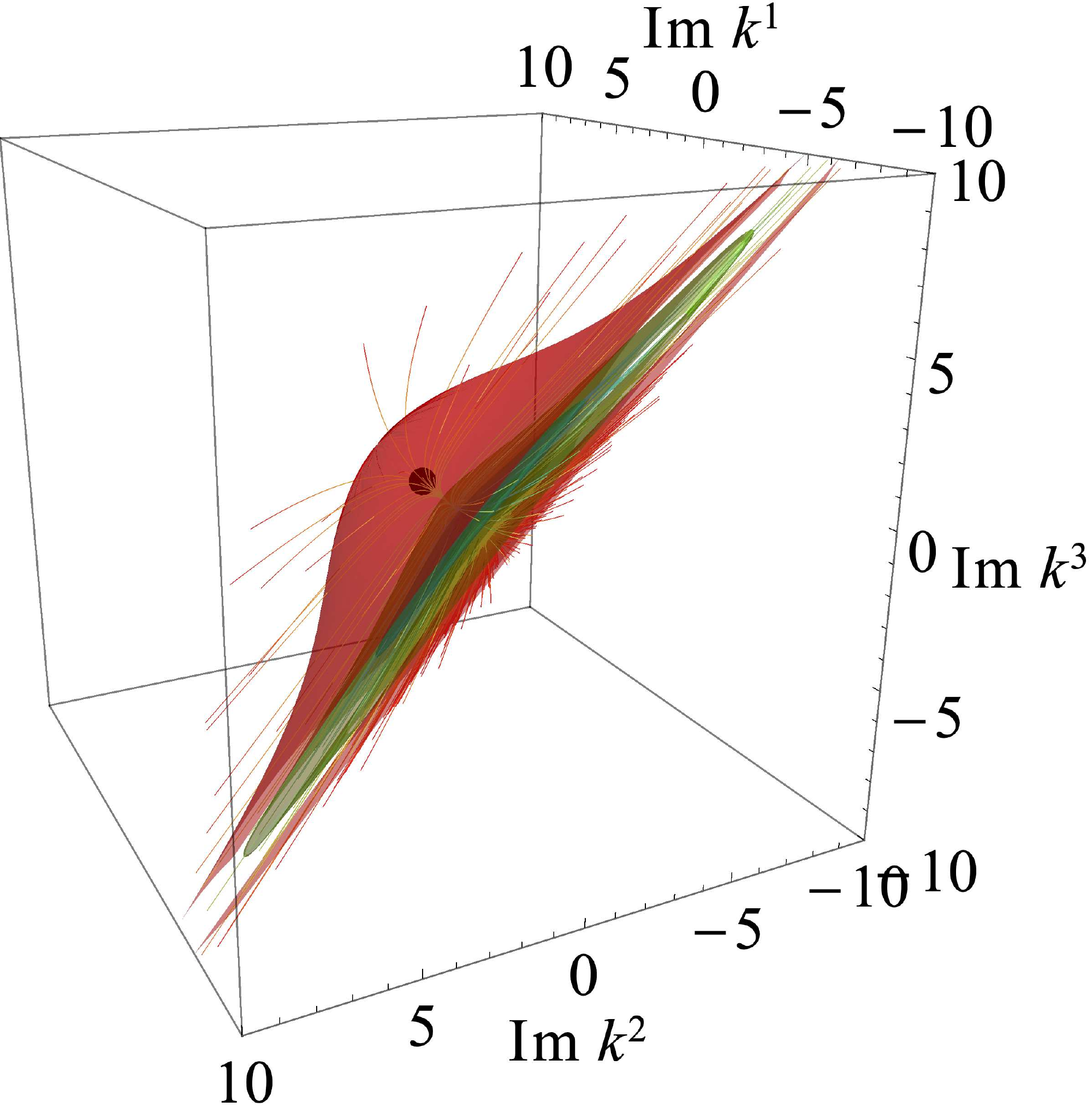}
    \includegraphics[width=0.46\linewidth]{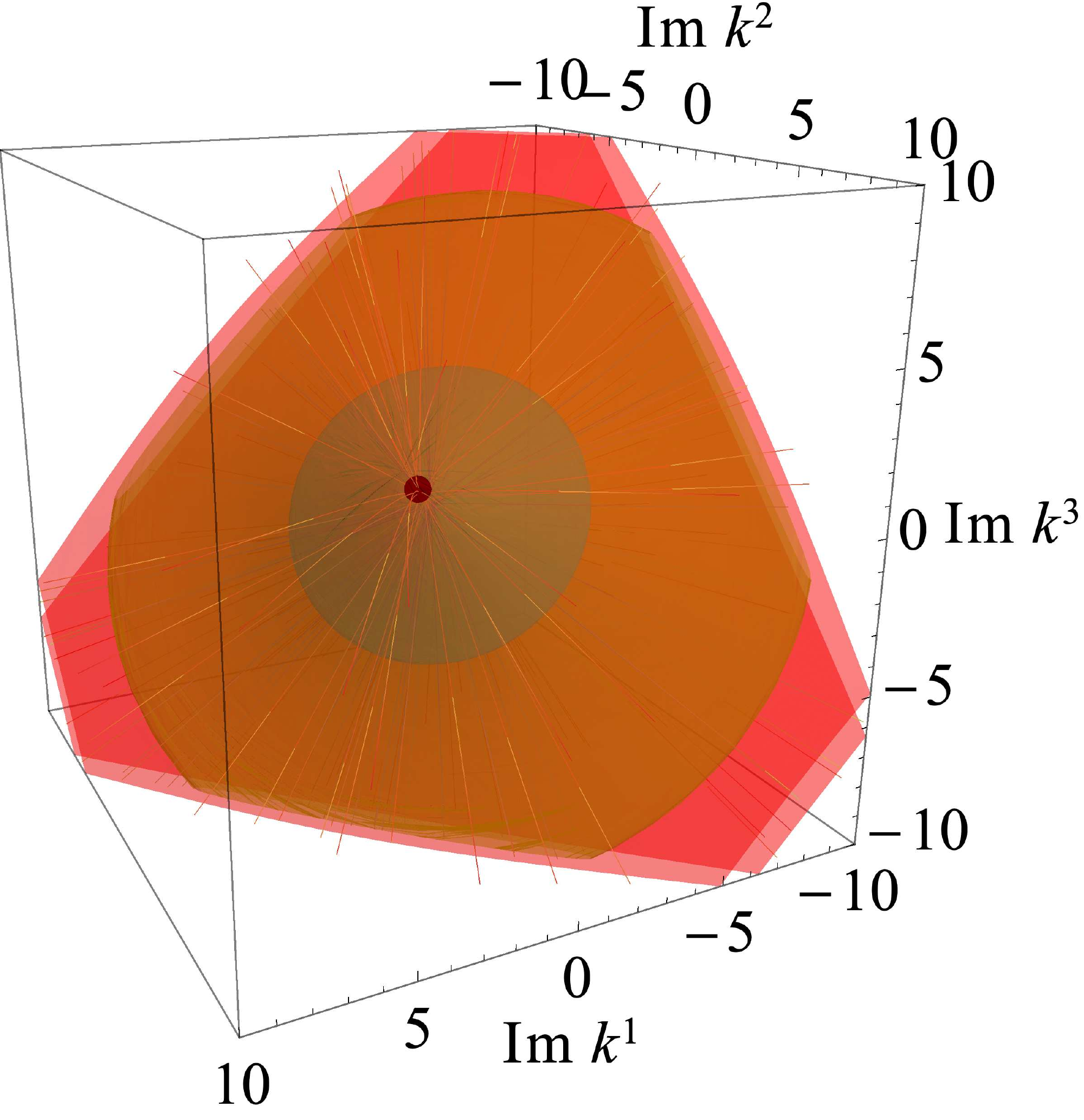}
    \caption{Same as Fig.~\ref{fig:DualThimble3D111} but for $\vec{v}=(0.5,0.5,0.5)$.}
    \label{fig:DualThimble3D050505}
\end{figure*}

\begin{figure*}[htb]       
    \includegraphics[width=0.46\linewidth]{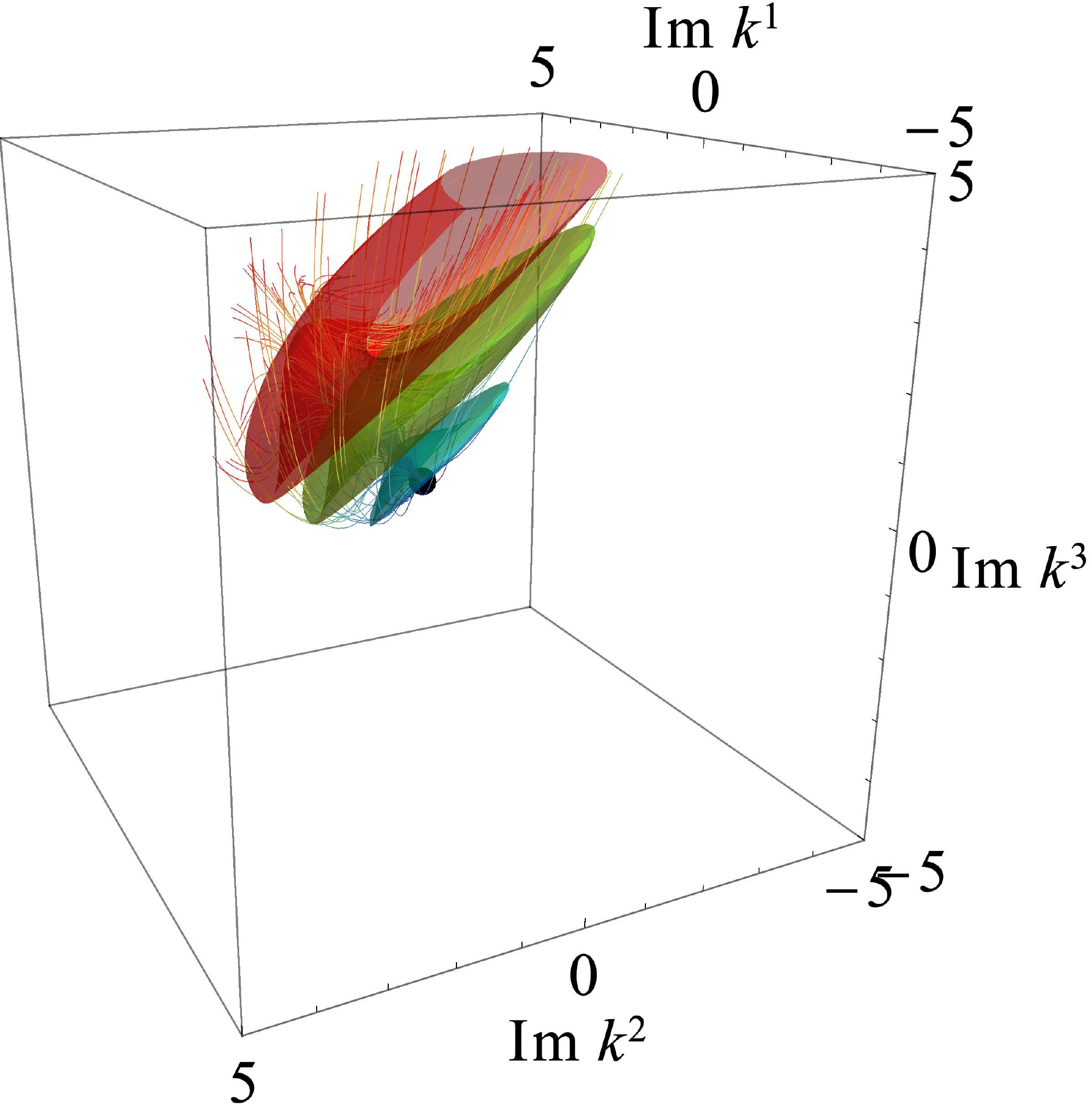}
    \includegraphics[width=0.46\linewidth]{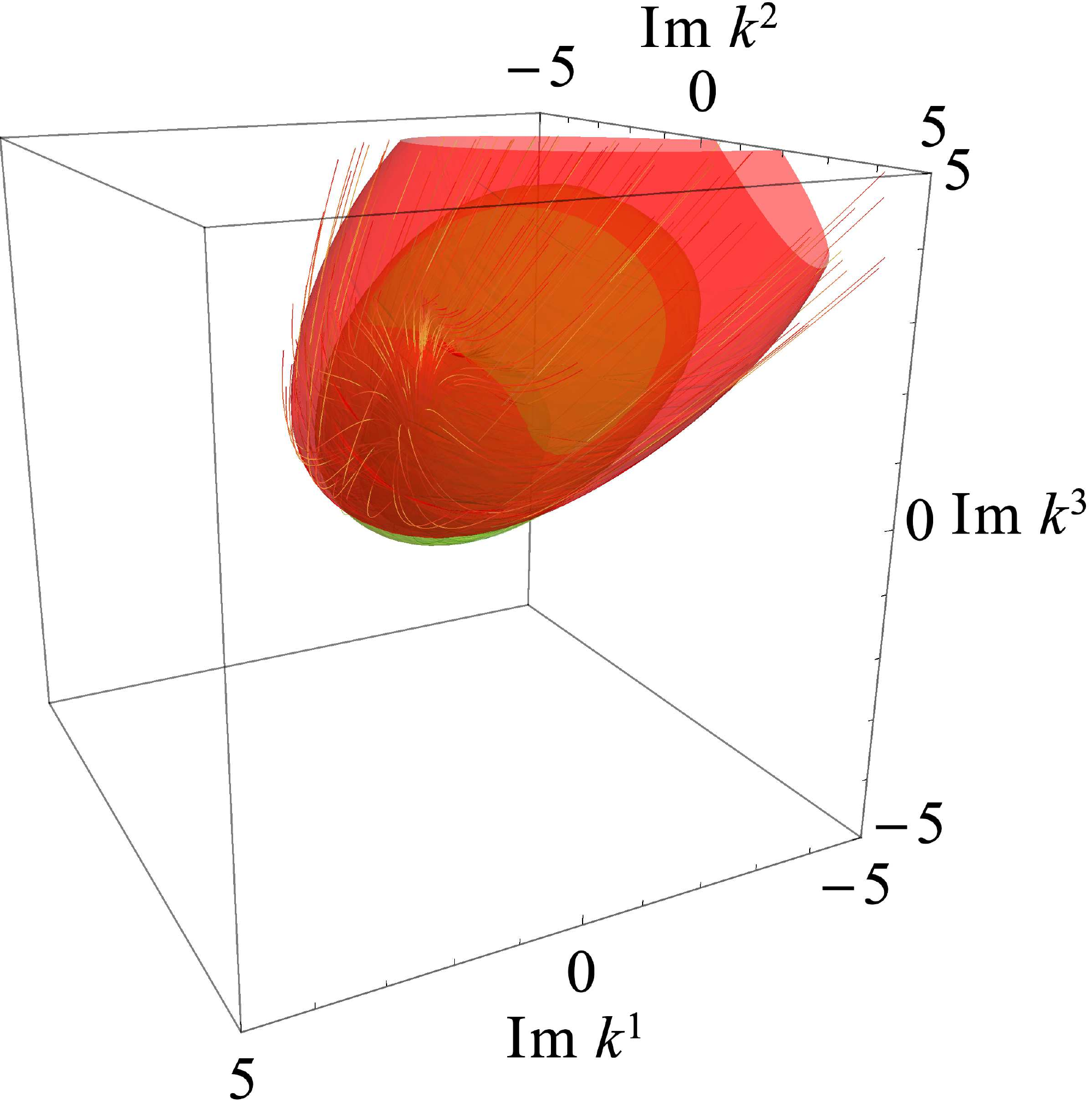}
    \caption{Same as Fig.~\ref{fig:DualThimble3D111} but for $\vec{v}=(0.5,0.25,0)$.}
    \label{fig:DualThimble3D050250}
\end{figure*}

\begin{figure*}[htb]       
    \includegraphics[width=0.46\linewidth]{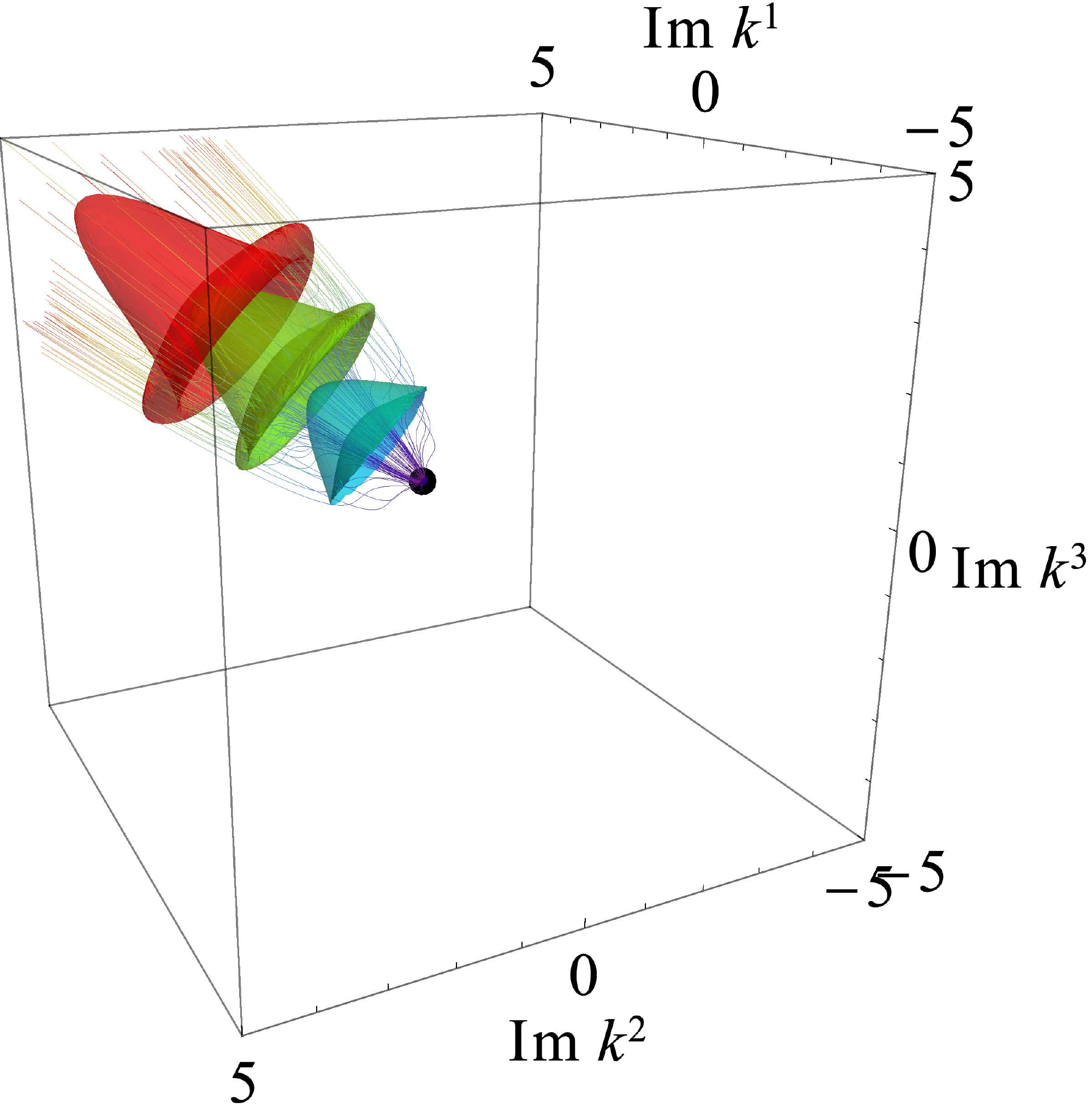}
    \includegraphics[width=0.46\linewidth]{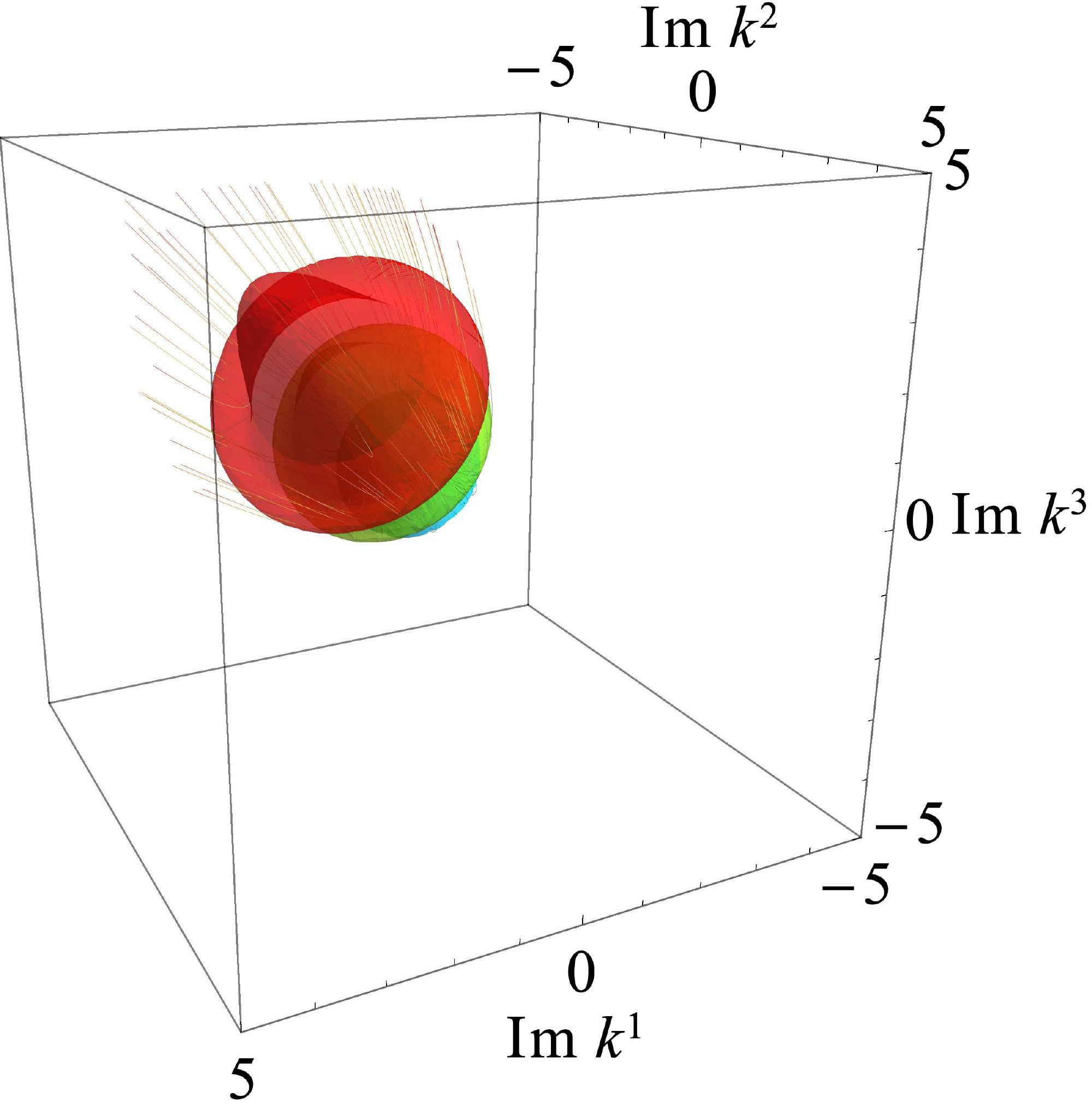}
    \caption{Same as Fig.~\ref{fig:DualThimble3D111} but for $\vec{v}=(0,0,0)$.}
    \label{fig:DualThimble3D000}
\end{figure*}

In Figs.~\ref{fig:DualThimble3D111}, \ref{fig:DualThimble3D050505}, \ref{fig:DualThimble3D050250} and \ref{fig:DualThimble3D000}, the dual thimble $\mathcal{K}_+$ attached to the highest critical point $k_+$ is shown for some different $\vec{v}$.
The dual thimbles are now 3-dimensional manifold. 
We hence show their sections at some different $s$, which are isomorphic to $S^2$, for better visibility.
Note that the projection of one of these sections onto the section at $k^0 = \re k^1 = \cdots = \re k^d$ can have self-intersections although it does not in the $(d+1)$-dimensional complex space (just like the Klein bottle);
this situation is indeed realized in Figs.~\ref{fig:DualThimble3D050250} and \ref{fig:DualThimble3D000}.
In Figs.~\ref{fig:DualThimble3D111} and \ref{fig:DualThimble3D050505}, we can verify that the origin is enclosed in the sections of $\mathcal{K}_+$ at sufficiently large $s$ and hence $\braket{\mathcal{C},\mathcal{K}_+} = \pm 1$.
On the other hand, Figs.~\ref{fig:DualThimble3D050250} and \ref{fig:DualThimble3D000} show that the origin is not enclosed in the sections, implying $\braket{\mathcal{C},\mathcal{K}_+} = 0$.
We can then confirm that $(v^1-1)^2+(v^2-1)^2+(v^3-1)^2<1$ is also the sufficient condition for instability for the DR given by Eq.~(\ref{eq:ToyDR3D}).


\section{Conclusion}
Based on the Lefschetz thimble method, we have reformulated linear instability analysis for field quantities that obey partial differential equations in an unlimited, not necessarily one-dimensional,  spatial domain. Instead of detecting coalescence of two complex roots of the DR as in the conventional theory, which is all but impossible in more than one spatial dimension in fact, we employ the intersection form defined between the original integration contour and the dual thimbles, which are obtained by solving the gradient flow equations. We have derived the explicit asymptotic formula for the retarded Green function that gives the evolution of perturbations in a wave packet form. Our new formulation is not only more mathematically rigorous than the classical method, properly picking up the critical points and associated Lefschetz thimble, or the steepest descent paths, that contribute to the asymptotic behavior, but is also practically useful mainly because we do not have to solve simultaneously the DR $\Delta(k)=0$ and some of its derivatives $\partial_i\Delta(k)=0$ explicitly and repeatedly for essentially all $k\in\mathbb{C}^{d+1}$ as in the conventional theory. Since the basic equation we assumed are quite generic, we believe that this formulation will have broad applications. As a matter of fact, we are currently applying the method to the analysis of collective neutrino oscillations, a nonlinear problem, which is notorious for its difficulties in direct numerical solutions and is our original motivation for this work. The results will be reported elsewhere soon. It may be also interesting to apply the method with some extensions possibly to the problem with oscillating sources.

\begin{acknowledgments}
T.M is supported by JSPS Grant-in-Aid for JSPS Fellows (No. 19J21244) from the Ministry of Education, Culture, Sports, Science and Technology (MEXT), Japan. This work is also supported by the Grant-in-Aid for the Scientific Research (No. 16H03986) and Waseda University Grant for Special Research Projects (Project number: 2018K-263).
\end{acknowledgments}

%

\end{document}